\def\preprint{1}		
\def\comment#1{}

\if\preprint1
	\documentclass[usenatbib]{mn2e}
        \usepackage{natbib}
	\usepackage{times}
	\usepackage{graphicx}
	\usepackage{color}
	\usepackage{xcolor}
	\usepackage{calc}
\else
	\documentstyle[astrop-bib,referee,times]{mn}
	\newcommand{\includegraphics}[1]{}
\fi

\def\oversim#1#2{\lower0.5pt\vbox{\baselineskip0pt \lineskip-0.5pt
     \ialign{$\mathsurround0pt #1\hfil##\hfil$\crcr#2\crcr\sim\crcr}}}

\title[]
{
Shutting down or powering up a (U)LIRG?  Merger components in distinctly different evolutionary states in IRAS~19115-2124 (The Bird)}
\author[]
       {Petri V\"ais\"anen$^{1,2}$, 
       Juha Reunanen$^3$, 
       Jari Kotilainen$^{3,4}$, 
       Seppo Mattila$^{4}$,  
\newauthor   
       Peter H.\ Johansson $^{5}$, 
       Rajin Ramphul$^{1,6}$,
       Cristina Romero-Ca\~{n}izales$^{7}$ \\
       $^1$South African Astronomical Observatory, P.O.Box 9, Observatory 7935, Cape Town, South Africa,\\	
       $^2$Southern African Large Telescope, P.O.Box 9, Observatory 7935, Cape Town, South Africa,\\	
      $^3$Finnish Centre for Astronomy with ESO (FINCA), University of Turku, FI-21500 Piikki\"o, Finland\\
      $^4$Tuorla Observatory, Department of Physics and Astronomy, University of Turku, FI-21500 Piikki\"o, Finland\\
      $^5$ Department of Physics, University of Helsinki, FI-00014 Helsinki, Finland \\
      $^6$University of Cape Town, Astronomy Department, Private Bag X3, Rondebosch 7701, South Africa\\
      $^7$ N\'ucleo de Astronom\'ia de la Facultad de Ingenier\'ia, Universidad Diego Portales, Santiago, Chile 
}
\pubyear{2017}

\begin{document}

\maketitle

\begin{abstract}

We present new SINFONI near-infrared integral field unit (IFU) spectroscopy and SALT optical long-slit spectroscopy characterising the history of a nearby merging luminous infrared galaxy, dubbed the Bird (IRAS19115-2114).  The NIR line-ratio maps of the IFU data-cubes and stellar population fitting of the SALT spectra now allow dating of the star formation (SF) over the triple system uncovered from our previous adaptive optics data.  The distinct components separate very clearly in a line-ratio diagnostic diagram.  An off-nuclear pure starburst dominates the current SF of the Bird with 60--70\% of the total, with a 4--7 Myr age, and signs of a fairly constant long-term star formation of the underlying stellar population.  The most massive nucleus, in contrast, is quenched with a starburst age of $>40$ Myr and shows hints of budding AGN activity. The secondary massive nucleus is at an intermediate stage.  The two major components have a population of older stars, consistent with a starburst triggered 1 Gyr ago in a first encounter.  The simplest explanation of the history is that of a triple merger, where the strongly star forming component has joined later.  We detect multiple gas flows in different phases. The Bird offers an opportunity to witness multiple stages of galaxy evolution in the same system; triggering as well as quenching of SF, and the early appearance of AGN activity.  It also serves as a cautionary note on interpretations of observations with lower spatial resolution and/or without infrared data. At high-redshift the system would look like a clumpy starburst with crucial pieces of its puzzle hidden, in danger of misinterpretations.

\end{abstract}


\section{Introduction}

Galaxies, by and large, come in two variants, blue and star-forming, and red and quiescent  \cite[e.g.][]{Kauffmann2003}.  Much of modern galaxy evolution study concentrates on understanding the processes involved in transforming galaxies to build up this bimodality.  This paper deals with feeding and quenching of star-formation, the evolutionary trajectory of a galaxy undergoing a radical transformation.

The strongest starburst activity in galaxies is linked to interactions or merging events \citep[e.g.][]{Sanders1996}, though secular evolution especially at higher redshift may also trigger strong starbursts \citep[e.g.][]{Shapiro2008}.  The most extreme cases are those of the ultra-luminous infrared galaxies (ULIRGs\footnote{ULIRGs are defined as $\log L/L_{\odot}>12.00$, and luminous IR galaxies, LIRGs, as $\log L/L_{\odot}=11.00-11.99$}) which invariably show perturbed morphologies in various stages of merging \citep{Sanders1996}.  During a merger, inflow of gas is believed to occur into the nuclear region, where a major starburst in a dusty environment would result, evident in a huge IR-luminosity elevating the system to an LIRG or ULIRG class \citep[e.g.][]{Barnes1996,Larson2016}.  Strong winds will develop, and with the growing central black hole and the following AGN activity these mechanisms will eventually sweep out the gas and terminate the starburst \citep[e.g.][]{Hopkins2006,Peter2009}. 

However, not all individual (U)LIRGs easily fit to the simple scenario above.  For example, the closest well-studied gas-rich merger, the Antennae,  has its strongest starburst in an overlap region between the merging nuclei \citep[e.g.][]{Mirabel1998,Karl2010}.  The details of the interaction clearly have an effect, and various more or less transient phases may be present, including off-nuclear star-bursts reported by e.g. \citet[][]{Jarrett2006, Vaisanen2008, Inami2010, Haan2011}, and a newly found case in IRAS~16516-0948 by Herrero-Illana et al. (2017, subm.). 
Moreover, the possibility of multiple interactions complicate the picture. This is a very realistic possibility due to (U)LIRGs naturally favouring galaxy group environments \citep[e.g.][]{Borne2000,Amram2007,Tekola2014}, though multiple components are often difficult to detect and quantify statistically because of the complex and dusty nature of the systems (though see \citet{Haan2011} who find a 6\% fraction for triple nuclei in the GOALS sample).

In this paper we present a case of a LIRG system, or essentially a ULIRG as it sits just below the arbitrary classification limit with $\log L/L_{\odot}=11.9$, which has all the ingredients of being a prototypical central-starburst dominated merger.  There are two massive nuclei in an advanced interaction with large tidal tails, with evidence of bars in the nuclear regions which are thought to facilitate the flow of gas into the central merging nuclei.  Our previous observations using SALT and VLT/NACO, reported in \citet{Vaisanen2008}, found a third kinematically separate component in the system, which Spitzer 24$\mu$m data at much lower spatial resolution suggested to be the source of the strong IR-luminosity detected by IRAS. 

We have taken new NIR IFU data and optical spectra of the system with the main goal of deriving the sequence of star formation in the  interacting system.  The IFU data allows both spatial and velocity-based disentangling of line emission, which is extremely important for complex targets such as LIRGs, where any integrated ratios are often misleading.  
Many advanced mergers are classified as "composite objects", i.e.\ assumed to be mixtures of AGN and star formation emission.  However, it has been shown that often shock emission together with SF mimics the composite signal \citep[e.g][]{Rich2014} without any real AGN contribution.  What is the case with the Bird?  Do the NIR line ratio diagnostics unambiguously show AGN emission, or can it also be due to SF induced shocks?  And if shocks, is it only shocks with no star formation?  In the latter case it would be a sign of quenching happening even though the target would be missed in typical post-starburst searches \citep[e.g.][]{Alatalo2016}, since even though SF has stopped, emission may still be seen due to shocks.


The connection of low-redshifts (U)LIRG studies to the broad cosmological view is through objects with similar IR luminosities being much more prevalent at high-$z$, and in fact dominating the star-formation density there \citep{PerezGonzalez2005}.  While the target of our study, IRAS~19115-2124 (see below), as a typical local (U)LIRG, lies well above the $z=0$ star-forming main-sequence (MS) of galaxies, with its stellar mass and star-formation rate (SFR) it would be located at the evolved MS at $z\sim2$ \citep{Whitaker2012}.   

The paper is organised as follows.  Section~{data} describes the observations and data reductions.  In Section~\ref{results} we present the new IFU data, with relevant feature maps, line fluxes and ratios, velocity fields and other derived quantities, and also present the results of stellar population and metallicity fitting to the optical spectra.  Section \ref{discussion} then discusses these results in the context of the overall evolutionary stage of the galaxy system, how they help to describe, together with the kinematics, the history and future of the (U)LIRG.  We end with a description of what the target would look like at higher redshift and lessons to be learned from that exercise.
We have adopted $H_0 = 73$ km s$^{-1}$ Mpc$^{-1}$, $\Omega_M = 0.27$, and $\Omega_{\Lambda} = 0.73$ cosmology (WMAP3) throughout.

\subsection{IRAS~19115-2124: the Bird}
\label{target}

\begin{figure}
\includegraphics[width=8.5cm,clip=true]{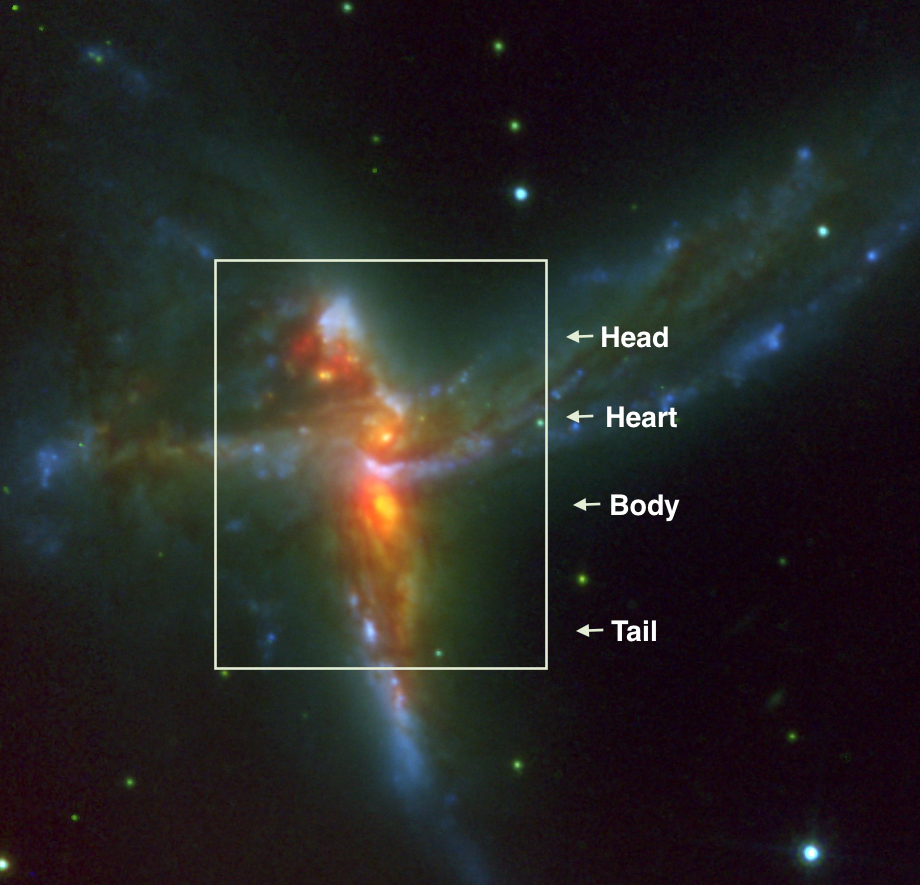}
\caption{\label{bird} Left:  A three-colour image of  IRAS 19115-2124, aka the Bird, made by combining  a K-band AO-image with HST I and B-bands (see V08).  The box indicates the position of the SINFONI K250 datacube, it is approximately $8 \times 10$ arcsec in size.  1\arcsec $\approx$ 1 kpc at the distance of the Bird.  The labels name the major features of the galaxy used throughout the paper. }
\end{figure}

The observed target, IRAS~19115-2124, also known as ESO~593-IG008, is at a distance of 200 Mpc ($z\approx0.05$).  Its infrared luminosity is $\log L / L_{\odot} = 11.9$ just shy of a ULIRG classification.  \citet[][henceforth V08]{Vaisanen2008} presented a detailed study of the system based on near-infrared (NIR) $K$-band adaptive optics (AO) imaging with $0.1\arcsec$ spatial resolution, optical long-slit spectroscopy, and archival optical HST and mid-infrared (MIR) Spitzer imaging.  The star formation rate (SFR) of the entire system was measured to be $\sim190$ M$_{\odot}$ yr$^{-1}$ from the far-IR SED.  The two main nuclei, as revealed by the NIR observations, are at a $\sim 1.5$ kpc projected (1.5\arcsec) distance from each other, and the system shows 20 kpc scale tidal tails.  A third component, argued to be a satellite galaxy, was also identified from the NIR images, projected 2 kpc North of the two main nuclei.  Because of its optical and NIR appearance, we dubbed IRAS~19115-2124 "the Bird" (Fig.~\ref{bird}).  The two primary nuclei in this context are called the Heart and the Body components, while the southern tidal feature is the Tail and the East and West tidal extensions the Wings.  The third component north-east of the Heart is the Head of the Bird.  

\section{Observations and data reductions}
\label{data}

\subsection{SINFONI}

Observations presented here were obtained in ESO service mode using the NIR IFU spectrograph SINFONI \citep{Eisenhauer2003} on VLT/UT4 in Adaptive Optics assisted mode.  Several sets of data were taken.  We used the 250 mas SINFONI plate scale in K-band (wavelength range of 1.95 to 2.45 $\mu$m, R$\sim$4000) giving an 8" $\times$ 8" field-of-view and had two slightly offset visits to cover the system from Head to Tail.  Due to jittering and offsets the final FOV is 9.5" $\times$ 10.5" corresponding to 10 kpc scale at the distance of the galaxy, and the reconstructed pixel scale is 0".125.  Observations were made on 28 and 30 September, 2010, in Object-Sky-Object sequences with 300 sec frames. The total effective exposure time in the overlapping area was 3600 sec.  A single J-band (1.10 to 1.40 $\mu$m, R$\sim$2000) observation was made in exactly the same manner on 30 August, 2010, resulting in 1800 sec total exposure time. 

In addition, we took K-band sets in the higher spatial resolution 100 mas SINFONI plate-scale covering a 3" $\times$ 3" area centred in between the Heart and Head components.  These data were obtained on the nights of September 12 and 24, 2010, with 900 sec individual exposures with a total exposure time of 3600 sec.  The seeing varied between 0.6" and 0.9" during all these natural guide star AO assisted  observations.  There are no obvious point sources in the science frames to check the PSF, but the most point-like sources in the final images have FWHM$\sim$0.4".  A spectrophotometric standard star was also observed after each science data set.

The data reductions were performed with the standard ESO pipeline ESOREX \citep{Freudling2013} which consists of dark, flat, detector linearity, geometric distortion corrections and wavelength calibrations using Neon and Argon arcs.  Sky was subtracted using the nearest sky frames.  The spectrophotometric standards taken were used to correct for remaining telluric features and to provide the initial spectral flux calibration.   Heliocentric corrections were also made to the spectra. 
Different visits in the K-band were spatially registered using the K-band NACO AO-images from V08 and then combined.  The final flux calibrations were done by comparing total fluxes obtained by convolving the spectra through the Ks filter in various apertures over the system with photometry from V08 in the K-band, and with 2MASS magnitudes in the J-band.  We estimate the fluxes to be absolutely accurate to within 20\%, while any fluxes relative to each other within the J or K-bands are significantly more accurate and limited by measurement errors depending on the strengths of the features.

\subsection{SALT}

Optical spectroscopy using the Robert Stobie Spectrograph \citep[RSS,][]{Burgh2003} at the Southern African Large Telescope \citep[SALT,][]{DOD2006}  was obtained on the nights of 30 July, 2011 and 1 April, 2012, of the target at spectral resolution R$\sim$1100 using the PG900 grating at two slightly different grating angles, respectively.  The data in total cover the redshifted spectral region starting blue-ward of [OII]$\lambda$3727 past the [SII]$\lambda$6717,6731doublet until approximately 7500\AA.  While the earlier SALT observations presented in V08 were taken with a higher resolution grating to study kinematics of the ionised and cool gas of the system, the data presented here allow fitting of stellar population models and determinations of extinction and metallicity across the galaxy. The first set used an exposure of 1200 sec in length, while the second had 1800 sec.  The observations were done at a single position angle of PA=10 degrees going though the Head and the Body+Tail, and covering the Heart just to the east of its nucleus (see Fig.~1 in V08 where PA=190 corresponds to the PA used in this work).  

\begin{figure}
\includegraphics[width=8.5cm,clip=true]{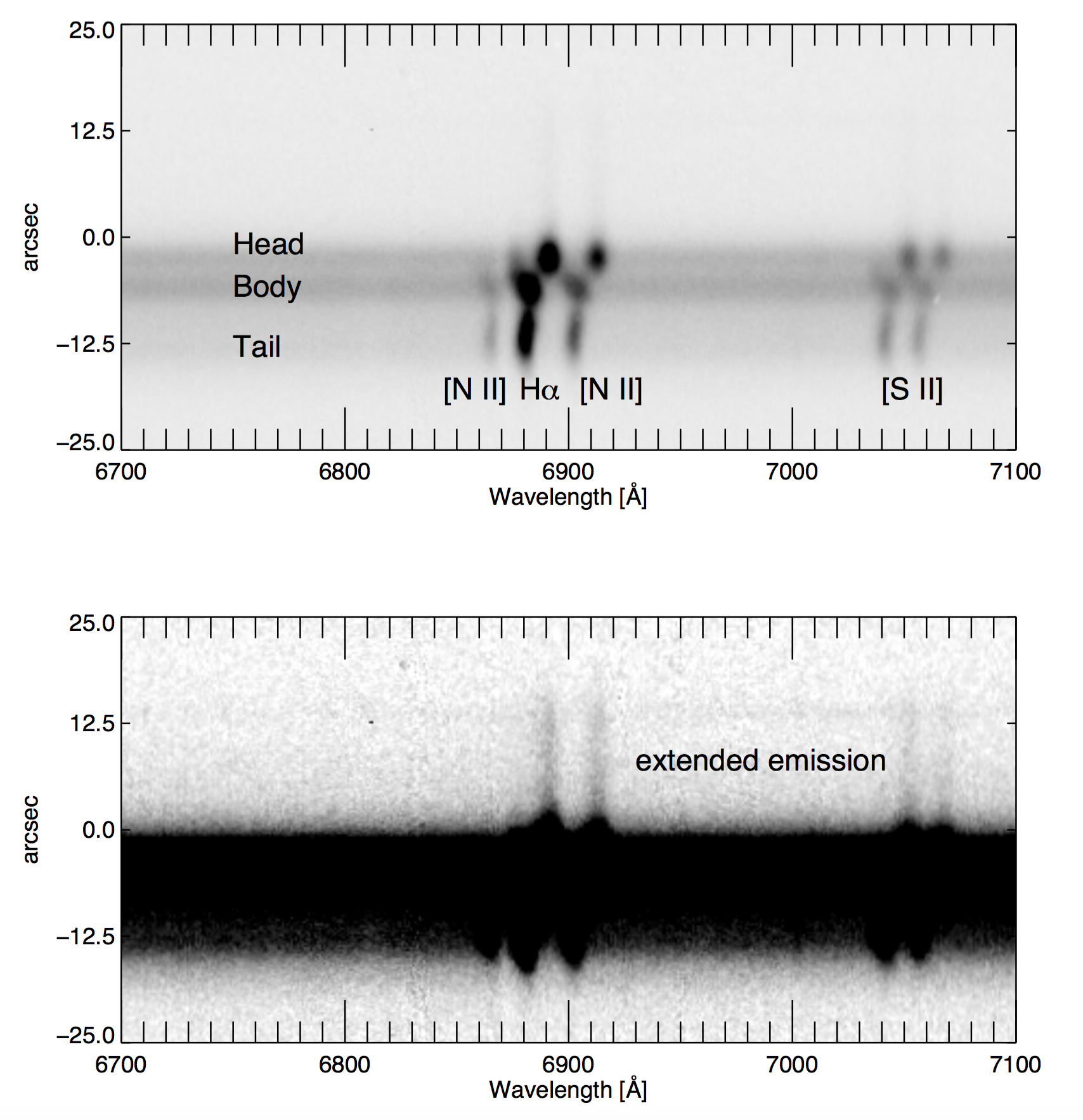}
\caption{\label{saltspec} A section of the SALT/RSS 2D spectrum around H$\alpha$.  Both panels show the same region, just scaled differently to highlight the extended diffuse ionised gas north of the Head in the lower panel.  
}
\end{figure}

The two spectra from the SALT data pipeline PySALT \citep{Crawford2010} products were reduced and calibrated using standard {\textsc IRAF} routines in the twodspec package, in the manner described in V08.  Four apertures were extracted from both, corresponding to the Head, Tail, and a combination of the Body and Heart components as the two could not be easily isolated at the $\sim 1.6$ \arcsec\ seeing-constrained spatial resolution, and also an aperture north of the Head showing ionised emission, but not continuum (see Fig.~\ref{saltspec}). The spectra from the two epochs, which mostly overlap in wavelength, were then averaged per galaxy component and aperture, providing a wider wavelength range than individually.

\subsection{VISIR}

We also obtained VLT/VISIR \citep{Lagage2004} mid-infrared imaging of the Bird in an attempt to assess the PAH emission distribution.
The data were taken on 16 and 19 May, 2009, in approximately 0.8\arcsec\ (visual) seeing conditions, with 
the PAH2\_2 (at 11.8 $\mu$m) filter corresponding to the redshifted 11.2 $\mu$m PAH- feature and using the NeII\_1 at 12.3 $\mu$m filter as the off-feature comparison.  The total exposure time of the chopped observation for each was approximately 1800 sec.
Reductions were done using ESOREX.  It turned out the S/N is not enough for a detailed study of the PAH fluxes over the galaxy system, but the general features of the Bird are recognisable.

\section{Results}
\label{results}

\subsection{Analysis of the IFU data}

\begin{figure}
\includegraphics[width=8.5cm,clip=true]{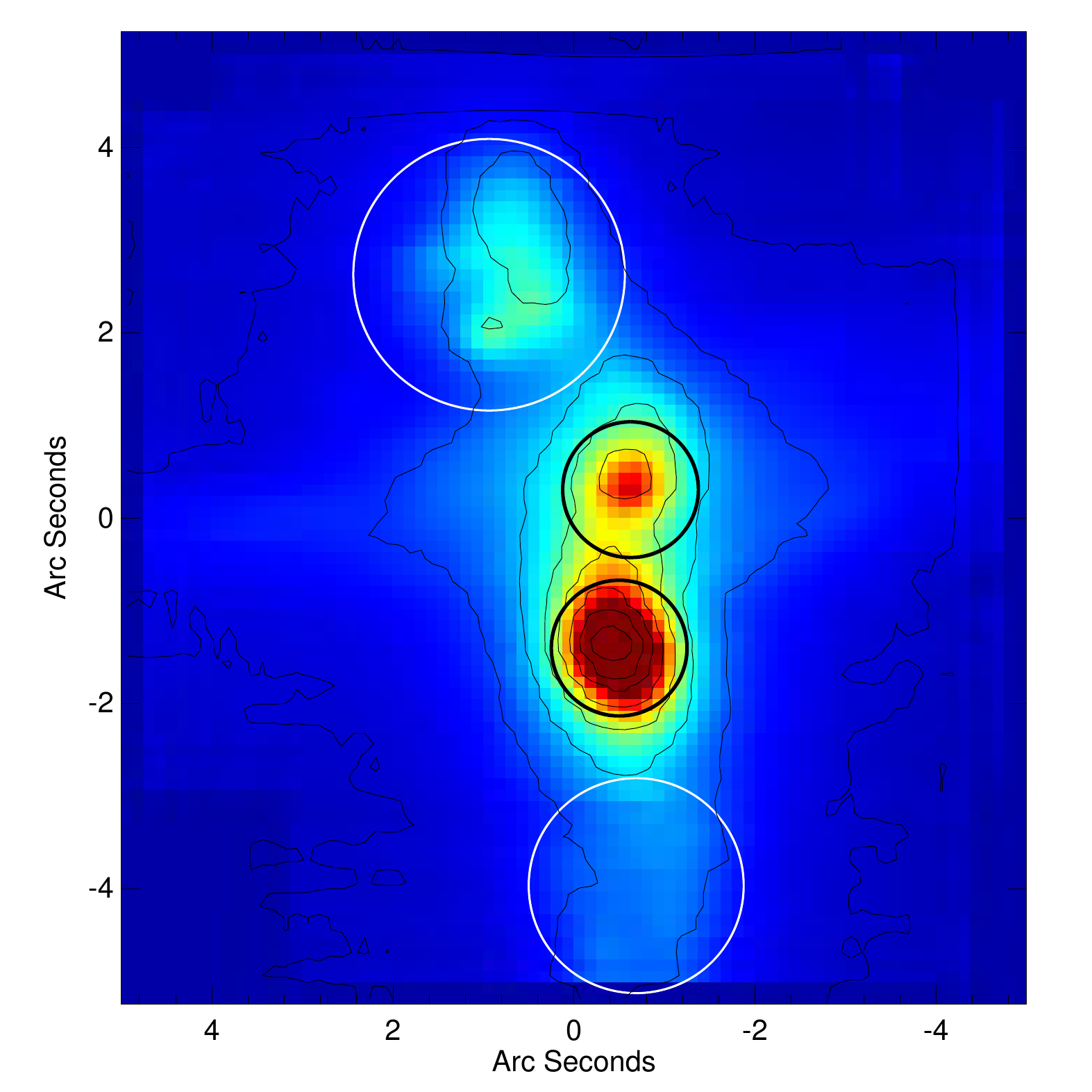}
\caption{\label{sinfoni_jk_imgs} The maps shows the collapsed K-band SINFONI data-cube with the corresponding J-band data-cube overlaid as contours.  The four circles show the apertures used to extract spectra of the relevant morphological components, Head, Heart, Body, and Tail, respectively from the top.  
}
\end{figure}

The SINFONI data-cubes provide a very rich data set for studying both the physics and kinematics of the system.   
The continuum subtracted line-maps described below were constructed using the QFitsView program\footnote{http://www.mpe.mpg.de/{\textasciitilde}ott/dpuser/qfitsview.html} and are summarised over all the velocity components of the system.
Spectra through the data-cubes were extracted using the same program inside the apertures listed in Table~\ref{apertures} and
shown in Fig.~\ref{sinfoni_jk_imgs}, overlaid on the collapsed K and J-band data-cubes showing the continuum light tracing the smooth older stellar light, but also potentially $\sim$10 Myr age red supergiant populations.   
Velocity fields were constructed using strong emission lines in the cubes by Gaussian fitting of the line in QFitsView.   
Below, we present the main results of the line-maps and the velocity fields in turn, before discussing the results in more detail in Section~\ref{discussion}.

\begin{table}
  \centering
  \begin{tabular}{lc}
  \hline
   \hline
Aperture     &      Radius  \\ 
     &  [arcsec and kpc] \\
\hline
 Body &   0.75  \\
 Body-nuc  & 0.25  \\
 Heart  & 0.75   \\
 Heart-nuc & 0.25  \\
 Head & 1.50  \\
 Tail &  1.25  \\
 \hline
\end{tabular}
  \caption{\small Sizes of the main apertures used in the analysis are listed, the four larger ones are also over-plotted in Fig.~\ref{sinfoni_jk_imgs}.  
  }
\label{apertures}
\end{table}

\begin{figure*}[h]
\includegraphics[width=7cm]{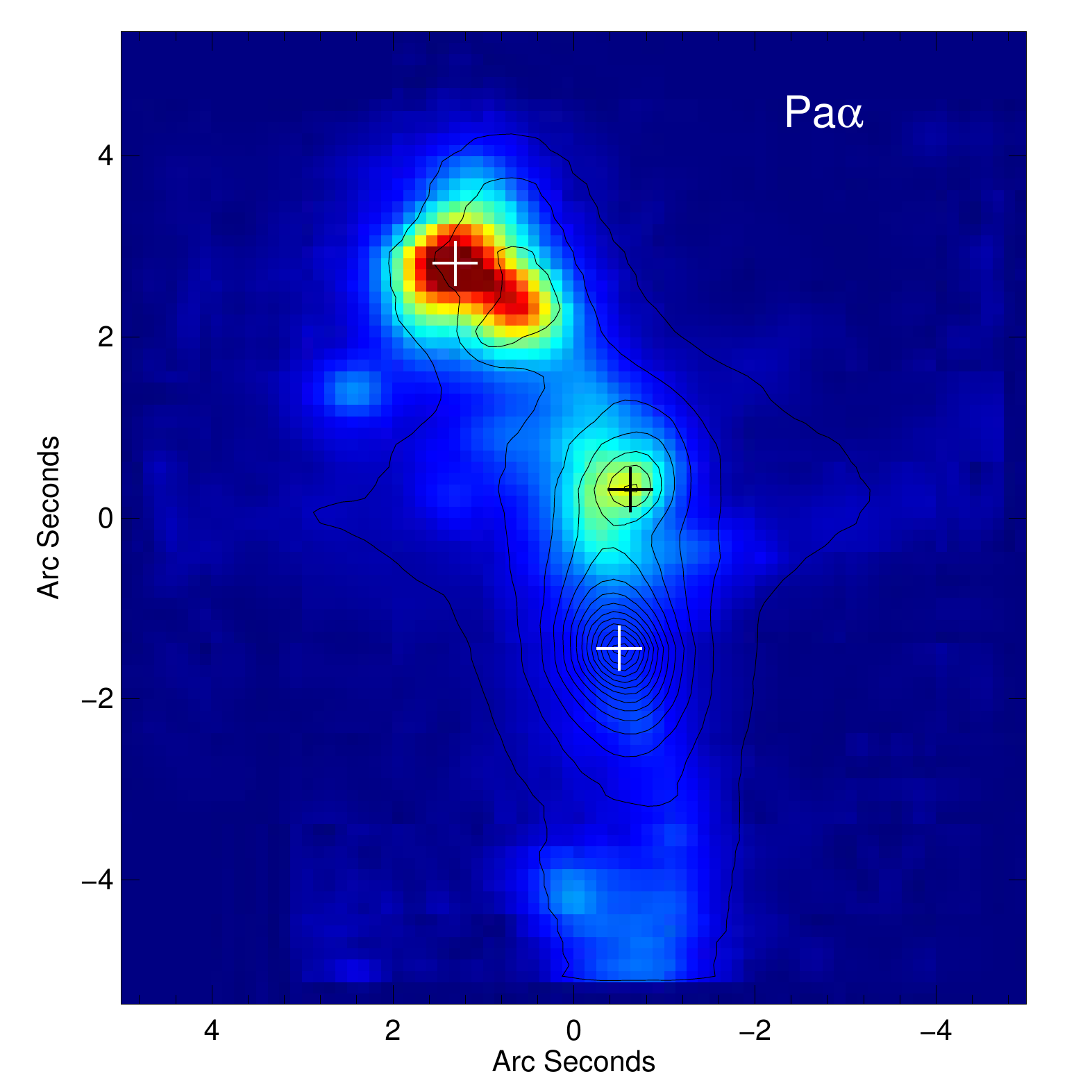}
\includegraphics[width=7cm]{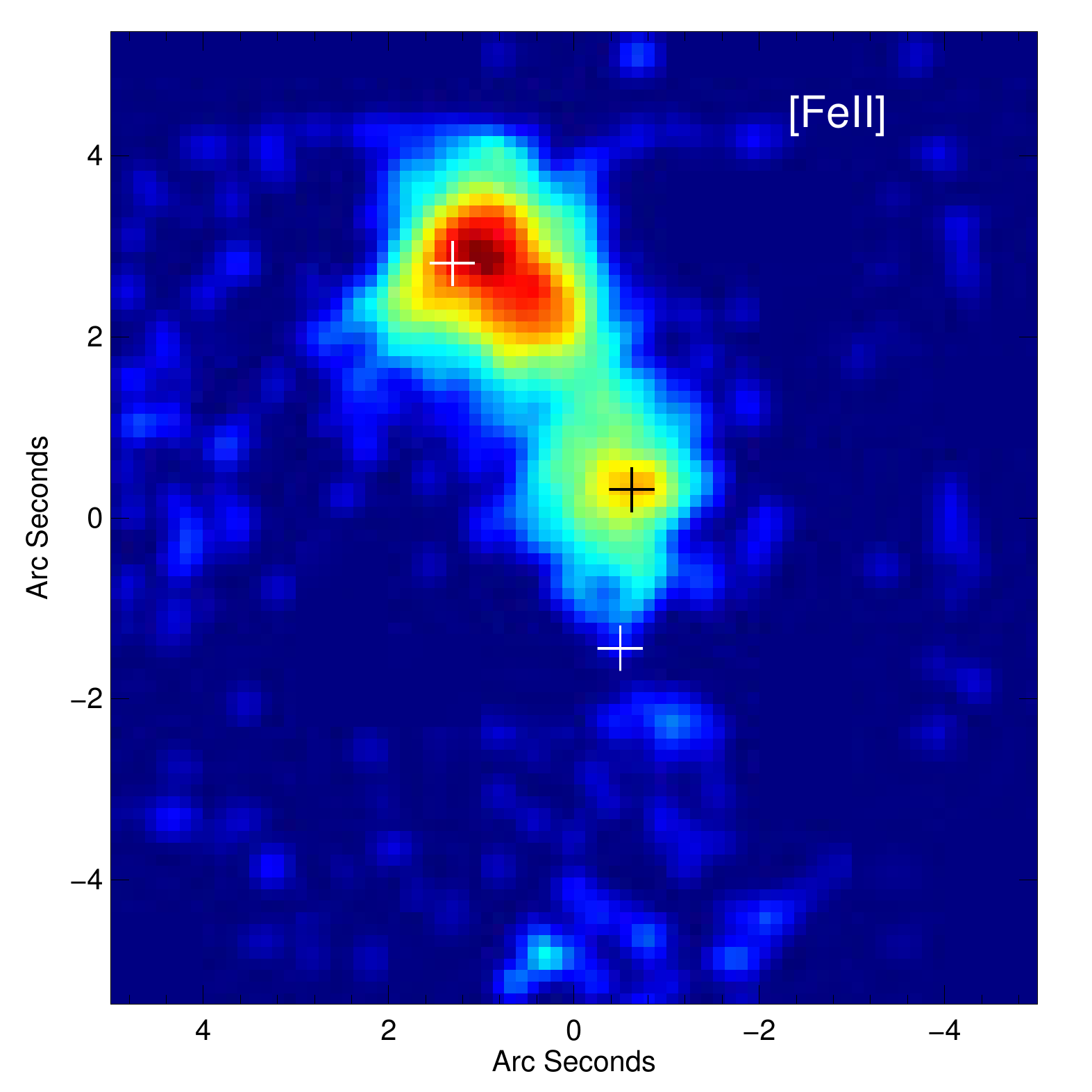}
\includegraphics[width=7cm]{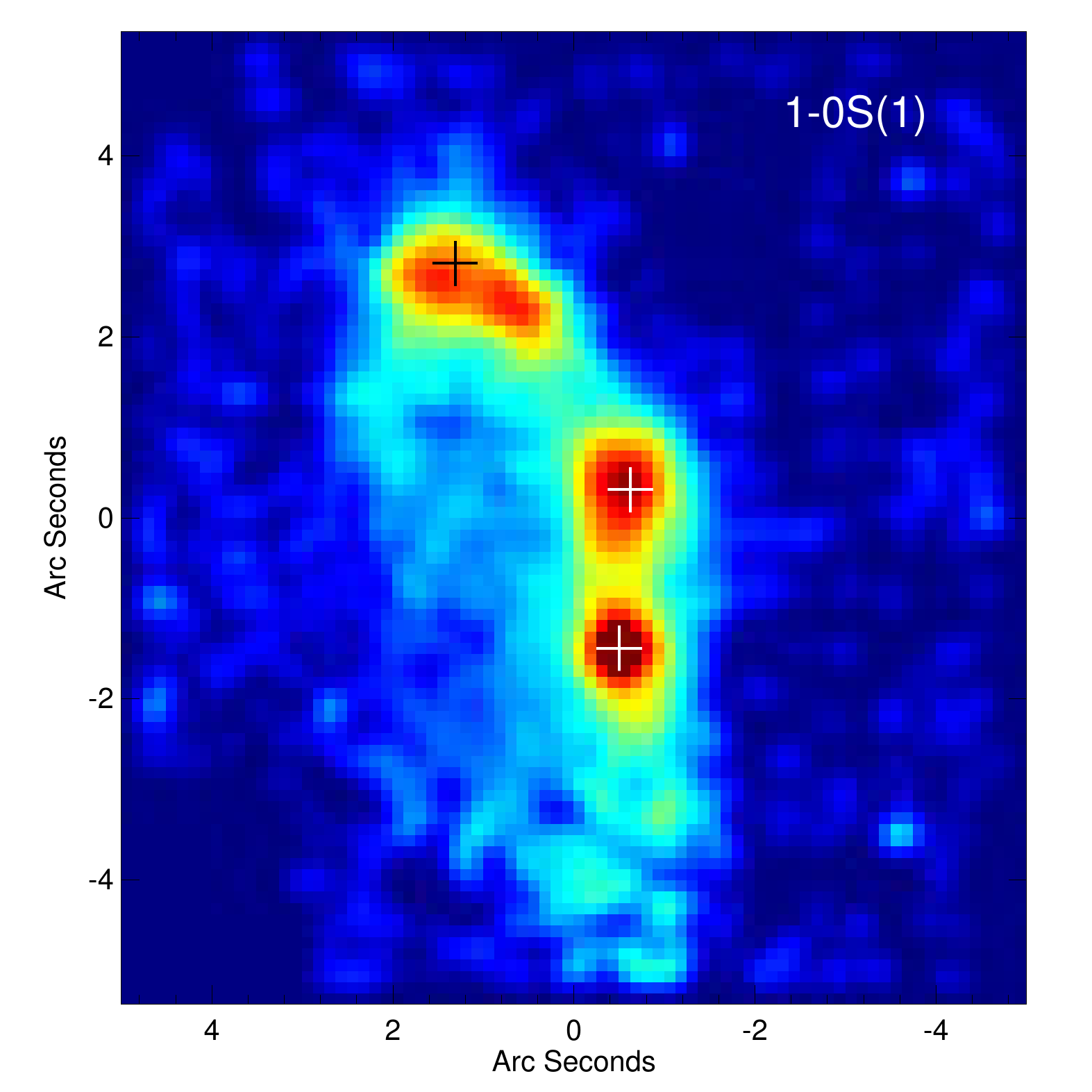}
\includegraphics[width=7cm]{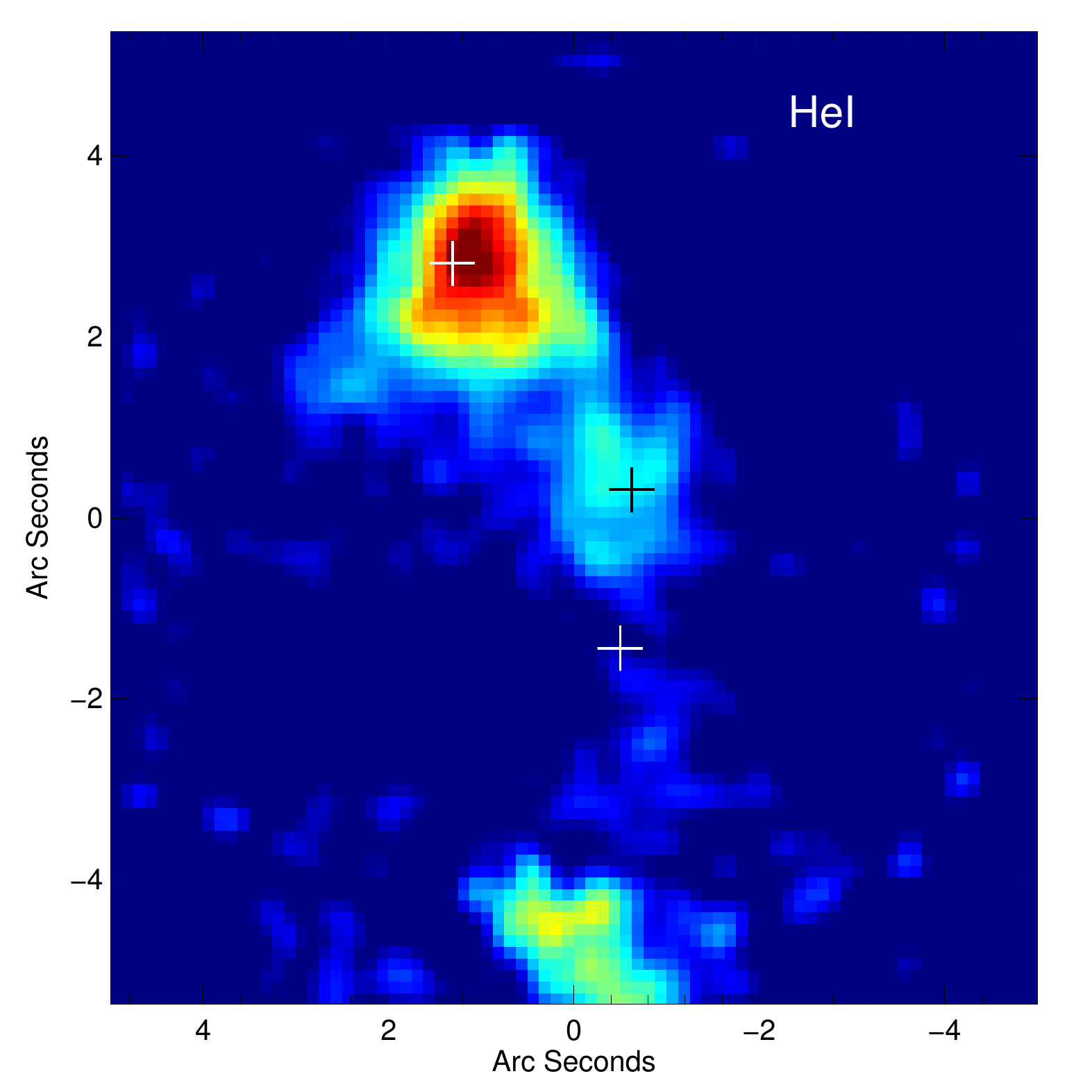}
\caption{\label{featuremaps} \small  Continuum subtracted line maps from the SINFONI data-cubes over all velocities.  The K-band continuum is shown as contours in the top left panel.  The crosses mark the locations, from the bottom, of the Body and Heart nuclei, and the peak of the Pa$\alpha$ flux in the Head component.  The maps show, clock-wise from the top left, Pa$\alpha$, [Fe II], He I 1.083$\mu$m, and H$_2$ (1-0) S(1).  Both [FeII] and HeI are more diffusely spread around the youngest regions of the Head and Heart than the more peaked Pa$\alpha$ emission.  The  molecular gas is concentrated in all the nuclei, and at the same time is also significantly more diffuse over the whole system than the other line emission.}
\end{figure*}

\subsection{Emission features}
\label{emission}

\begin{figure*}
\includegraphics[width=12cm,clip=true]{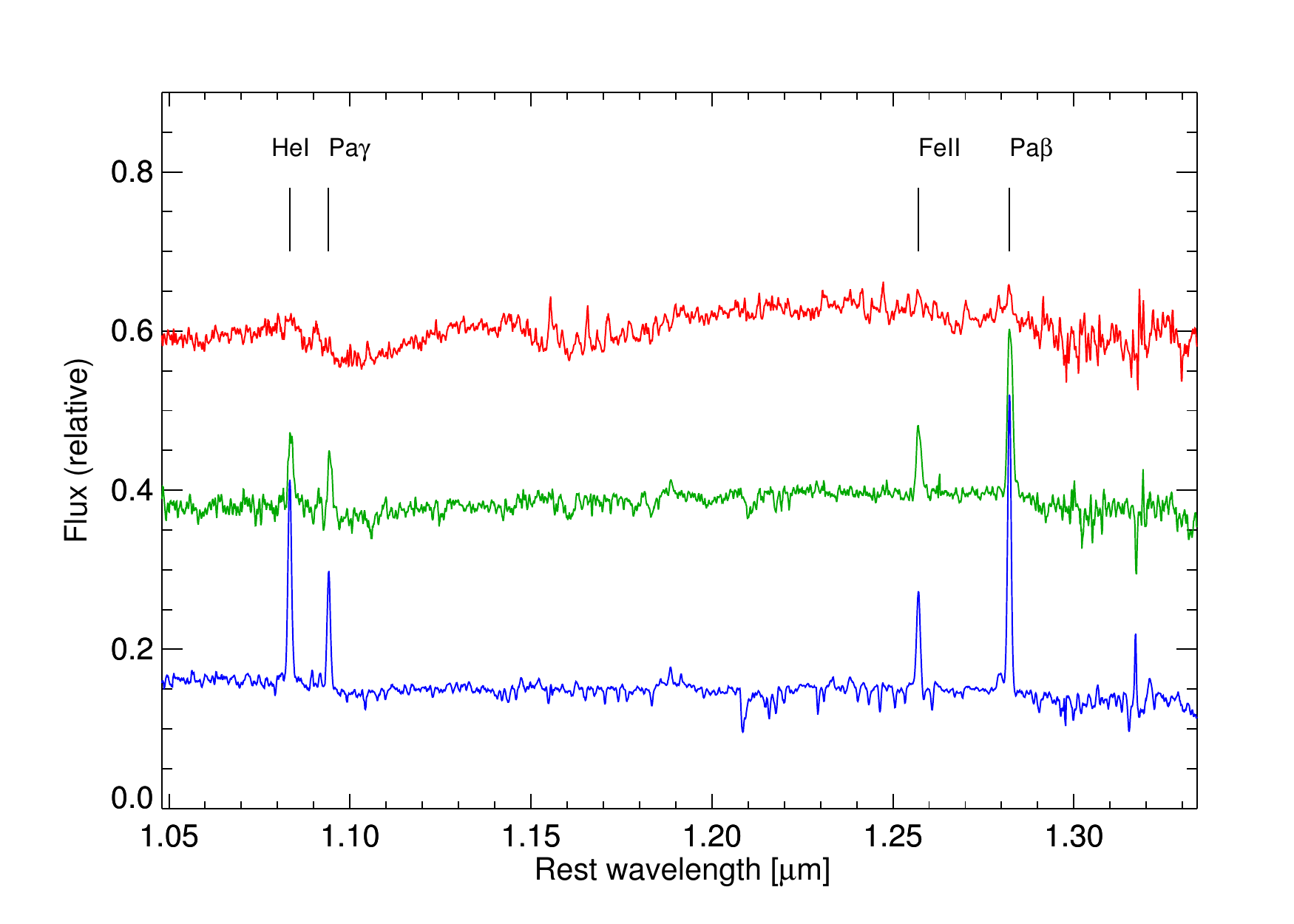}
\includegraphics[width=12cm,clip=true]{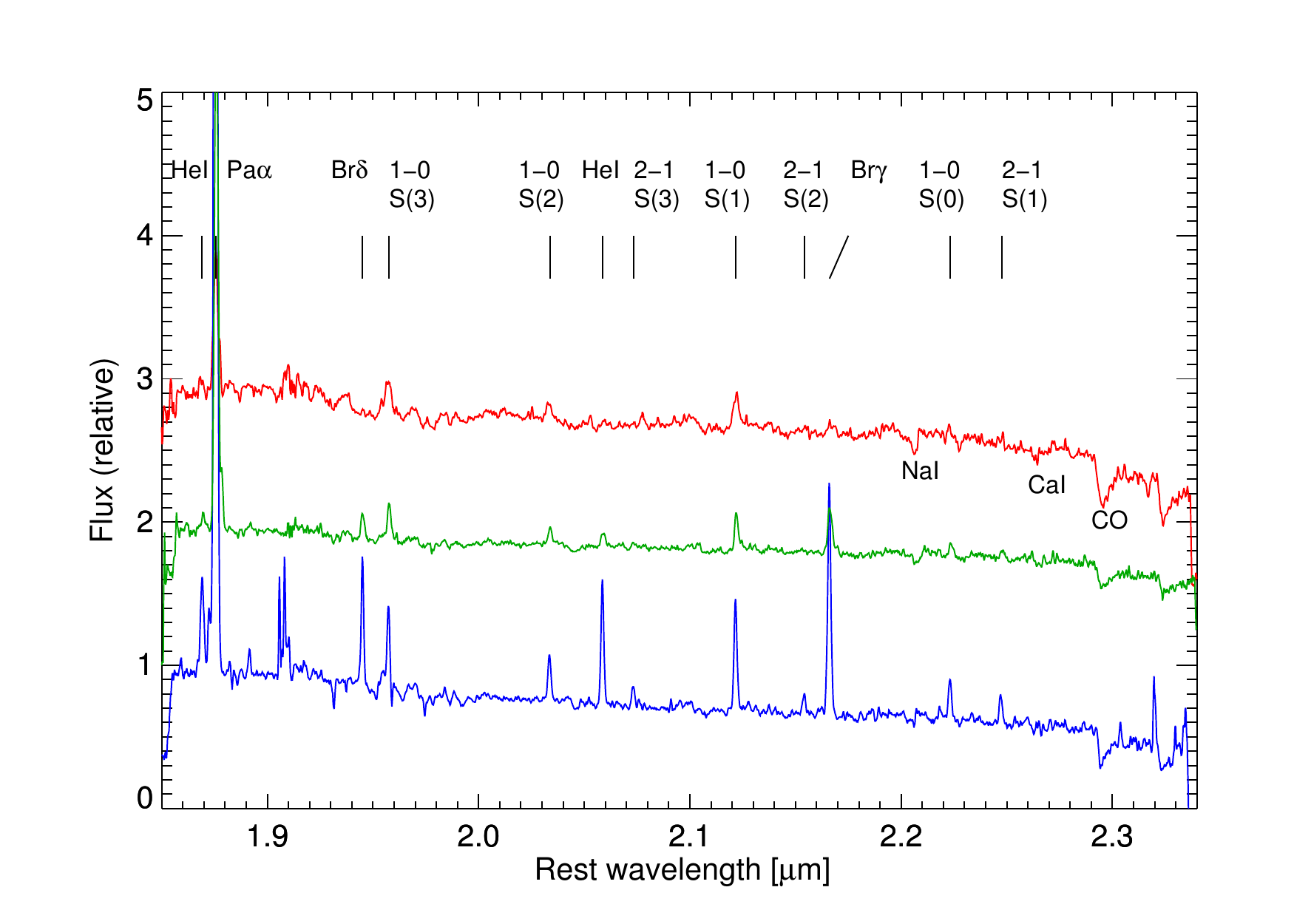}
\caption{\label{sinfoni_jk_specs}  Extracted 1D SINFONI spectra, J-band at the top and K-band at the bottom, for the small Body (red) and Heart (green) apertures, and for the Head (blue) showing the strongest emission lines.  The Tail is not plotted, it is quite similar to the Heart.  Main spectral features are indicated; there are many unmarked weaker H$_2$ lines in the J-band.    
}
\end{figure*}

Feature maps constructed from strong emission lines are shown in Figure~\ref{featuremaps}. These can be compared to the continuum image in Fig.~\ref{sinfoni_jk_imgs}.  
Ionised hydrogen in Pa$\alpha$ (the strongest of the recombination lines in our data, though several others also feature prominently, such as Pa\, $\beta$ and Br $\gamma$) in starburst galaxies traces the ionising young OB-star population of $\sim 10$ Myr or less.  [Fe II] in the J-band is thought to originate from shocking in supernova remnants and thus traces a slightly older population of $\sim 30$ Myr.  Potentially the youngest stellar population is seen in the neutral He I.   Especially the K-band data feature numerous lines from vibrational states of warm molecular hydrogen H$_2$. The strongest of these lines is the 1-0 S(1) transition at 2.1218 $\mu$m.  While the Pa \, $\alpha$, [Fe II], and He I trace fairly well the stellar light (though not with same relative strengths), H$_2$ emission is more extended, showing in areas with no obvious stellar light, such as in the eastern parts of the Bird, below its wing.  Representative extracted 1D spectra are shown in Fig.~\ref{sinfoni_jk_specs} highlighting the significant difference of spectra at the Body location compared to the others.

Emission line fluxes were calculated from emission feature maps, such as shown above, using IDL aperture photometry, and also from the spectra extracted from spatial apertures in the data-cubes. IRAF/splot routines were used in the latter case.  Since the continuum fitting method is different in the two methods, it serves as a check of consistency in case of the weaker lines where uncertainties are larger. 
The results are listed in Table~\ref{emissionlines}.  The listed errors include both direct measurement errors and the systematic ones including the differences in the fitting methods.  
Note that we did not attempt a stellar continuum subtraction, and thus the weakest recombination lines especially are lower limits due to underlying absorption lines contributing negatively to the fluxes -- no major results hinge on this uncertainly, however. 
An additional $\sim 20$\% uncertainty due to absolute flux calibration is not reflected, though this would not affect line ratios.

\begin{table*}
  \centering
  \begin{tabular}{llrrrrrrrr}
  \hline
   \hline
Regions       &                               &      Body &  Body-nuc  & Heart  & Heart-nuc-nuc & Head & Tail &  Integ.  \\
aperture       &                               &     0.75 &  0.25 &  0.75  &  0.25 & 1.5  & 1.25  &  full \\
 \hline
Line               & $\lambda_0$ [$\mu$m]   &      &     &    &     &    &   & \\
 \hline
   HeI(J)   &   1.0833   &    5.77 $\pm$  2.02  &    0.73 $\pm$  0.71  &   13.46 $\pm$  3.09  &    1.71 $\pm$  1.10  &   70.60 $\pm$  7.06  &   27.54 $\pm$  4.41  &  212.72 $\pm$ 12.26 & \\
      Pa$\gamma$   &   1.0941   &    4.72 $\pm$  2.39  &    0.56 $\pm$  0.82  &    9.15 $\pm$  2.75  &    1.24 $\pm$  0.94  &   34.24 $\pm$  5.97  &   12.60 $\pm$  3.81  &  114.12 $\pm$ 11.84 & \\
     FeII   &   1.2570   &    2.38 $\pm$  1.27  &    0.26 $\pm$  0.42  &    9.51 $\pm$  2.55  &    1.47 $\pm$  1.00  &   37.87 $\pm$  5.09  &    6.06 $\pm$  2.04  &  110.79 $\pm$  8.71 & \\
      Pa$\beta$   &   1.2822   &   10.13 $\pm$  2.41  &    1.69 $\pm$  0.83  &   24.13 $\pm$  3.96  &    3.64 $\pm$  1.53  &   96.75 $\pm$  8.02  &   26.42 $\pm$  4.17  &  248.82 $\pm$ 13.06 & \\
   \hline
      Pa$\alpha$   &   1.8756   &   55.45 $\pm$  4.43  &    6.44 $\pm$  1.51  &  127.95 $\pm$  6.74  &   20.92 $\pm$  2.72  &  495.55 $\pm$ 13.26  &  135.78 $\pm$  6.94  & 1286.40 $\pm$ 21.36 & \\
      Br$\delta$   &   1.9451   &    1.66 $\pm$  1.05  &    0.26 $\pm$  0.33  &    4.98 $\pm$  1.22  &    0.82 $\pm$  0.49  &   28.10 $\pm$  2.89  &    6.95 $\pm$  1.44  &   49.64 $\pm$  3.84 & \\
    1-0S(3)   &   1.9576   &   16.42 $\pm$  2.25  &    2.91 $\pm$  0.95  &   11.76 $\pm$  1.92  &    1.76 $\pm$  0.78  &   22.47 $\pm$  2.85  &   10.27 $\pm$  1.86  &  142.52 $\pm$  6.89 & \\
    1-0S(2)   &   2.0338   &    3.95 $\pm$  0.98  &    0.79 $\pm$  0.42  &    3.98 $\pm$  1.19  &    0.74 $\pm$  0.49  &    9.36 $\pm$  1.77  &    4.56 $\pm$  1.08  &   41.91 $\pm$  3.73 & \\
  HeI(K)   &   2.0587   &    1.35 $\pm$  0.69  &    0.04 $\pm$  0.13  &    3.57 $\pm$  1.12  &    0.57 $\pm$  0.45  &   23.25 $\pm$  2.87  &    5.25 $\pm$  1.36  &   45.32 $\pm$  4.01 & \\
  2-1S(3)   &   2.0735   &    0.00 $\pm$  0.00  &    0.00 $\pm$  0.00  &    0.00 $\pm$  0.00  &    0.00 $\pm$  0.00  &    1.91 $\pm$  0.17  &    0.00 $\pm$  0.00  &   0.00 $\pm$ 0.00 & \\
  1-0S(1)   &   2.1218   &   10.04 $\pm$  1.86  &    1.88 $\pm$  0.83  &   10.18 $\pm$  1.87  &    1.51 $\pm$  0.74  &   23.20 $\pm$  2.85  &   10.82 $\pm$  1.96  &  109.78 $\pm$  6.24 & \\
  2-1S(2)   &   2.1542   &    0.00 $\pm$  0.00  &    0.00 $\pm$  0.00  &    0.00 $\pm$  0.00  &    0.00 $\pm$  0.00  &    1.80 $\pm$  0.20  &    0.00 $\pm$  0.00  &   0.00 $\pm$ 0.00 & \\
      Br$\gamma$   &   2.1661   &    7.22 $\pm$  1.68  &    0.91 $\pm$  0.59  &   13.92 $\pm$  2.33  &    2.38 $\pm$  0.96  &   54.57 $\pm$  4.61  &   14.04 $\pm$  2.34  &  142.93 $\pm$  7.46 & \\
  1-0S(0)   &   2.2233   &    3.17 $\pm$  1.14  &    0.58 $\pm$  0.52  &    3.51 $\pm$  1.09  &    0.49 $\pm$  0.39  &   10.91 $\pm$  2.15  &    4.35 $\pm$  1.28  &   48.32 $\pm$  4.31 & \\
  2-1S(1)   &   2.2477   &    4.82 $\pm$  1.27  &    0.66 $\pm$  0.50  &    3.42 $\pm$  1.01  &    0.40 $\pm$  0.36  &    7.55 $\pm$  1.56  &    3.47 $\pm$  0.99  &   41.94 $\pm$  3.63 & \\
          \hline
\end{tabular}
  \caption{\small  Observed line fluxes, in units of $10^{-16} \, erg \, s^{-1} \, cm^{-2}$.  Zero values mark non-detections.  Aperture size is in units of kpc (and arcsec).    
 }
\label{emissionlines}
\end{table*}

Using the measurements of the line fluxes, we tabulate relevant line ratios in Table~\ref{lineratios}.  The recombination line ratios were also used to calculate extinction using the Calzetti reddening law for a foreground dust screen. 

\begin{table*}
  \centering
  \begin{tabular}{lccccccc}
  \hline
   \hline
Region   &                              Body & Body-nuc & Heart  & Heart-nuc & Head &  Tail  & Integ. \\
\hline
FeII / Pa$\beta$                              &   0.23  &  0.15  &  0.39  &  0.40 &   0.39  &  0.23  &  0.45 \\
1-0 S(1) / Br$\gamma$                                     &  1.39    &  2.07  &  0.73  &  0.63  &  0.43  &  0.77  &  0.77 \\
1-0 S(1) / Pa$\alpha$                                    &  0.18   & 0.29  &  0.08   & 0.07   & 0.05   & 0.08   & 0.09 \\
FeII / 1-0 S(1)                                  &   0.24   & 0.14  &  0.93   & 0.97   & 1.63  &  0.56   & 1.01 \\
HeI / 1-0 S(1)                                  &   0.13   & 0.02  &  0.35  &  0.38  &  1.00  &  0.49  &  0.41 \\            
1-0 S(0) / 1-0 S(1)                       &  0.32  &  0.31  &  0.34   & 0.32   & 0.47  &  0.40 &   0.44  \\
1-0 S(2) / 1-0 S(1)                         & 0.39   & 0.42  &  0.39   & 0.49  &  0.40  &  0.42  &  0.38 \\
1-0 S(2) / 1-0 S(0)                         & 1.25   & 1.36  &  1.13   & 1.51  &  0.86  &  1.05  &  0.87 \\
1-0 S(3) / 1-0 S(1)                          & 1.64   & 1.55   & 1.16  &  1.17  &  0.97  &  0.95 &   1.30 \\
 2-1 S(1) / 1-0 S(1)                          & 0.48  &  0.35   & 0.34  &  0.26  &  0.33   & 0.32  &  0.38 \\
 HeI /  Br$\gamma$                     &   0.29  &  0.07   &  0.39   &  0.46  &  0.68 &  0.42 & 0.40 \\
\hline
$A_V$ (Pa$\beta$/Pa$\gamma$)   &   2.1  &  6.2  &  4.6   &  5.9  &  5.4 &  1.8 &  2.3 \\
$A_V$ (Pa$\alpha$/Pa$\beta$)    &   6.4  &  4.0  &  6.2   &  6.7  &  6.0 &  6.0 & 6.0 \\
$A_V$ (Br$\gamma$/Pa$\alpha$) &   10.2  &  12.0  &  6.1   &  7.1  &  6.4 &  5.0 & 6.6  \\
$A_V$ (Br$\gamma$/Br$\delta$)  [weak] &   32.3  &  25.6  &  18.7  &  19.9  &  7.5 &  8.7 & 19.6  \\
 \hline
\end{tabular}
  \caption{\small Column densities and line ratios.  Note that the ratios are not extinction corrected in the table, though all figures and maps showing line ratios are corrected for extinction adopting the values using the Br$\gamma$/Pa$\alpha$ derived  $A_V$.  
 }
\label{lineratios}
\end{table*}

\subsection{Kinematics and masses}
\label{kinres}

First, we fit Gaussians to the strong lines in the 1D extracted spectra (Fig.~\ref{sinfoni_jk_specs}) of the various apertures to obtain line-of-sight heliocentric radial velocities and velocity dispersions.  Results are listed per aperture separately for ionised and molecular gas in Table~\ref{mainkinematics}.  We also add H$\alpha$ line values from SALT long-slit data.  The radial velocities of the latter are consistent with the corresponding NIR values with one notable exception:  in the Body, the most dust obscured region of the system, the optical emission picks up a different velocity signal, which most likely is related the Heart spiral arms, rather than the deeper component probed by the SINFONI data. 

\begin{figure}
\includegraphics[width=8cm,clip=true]{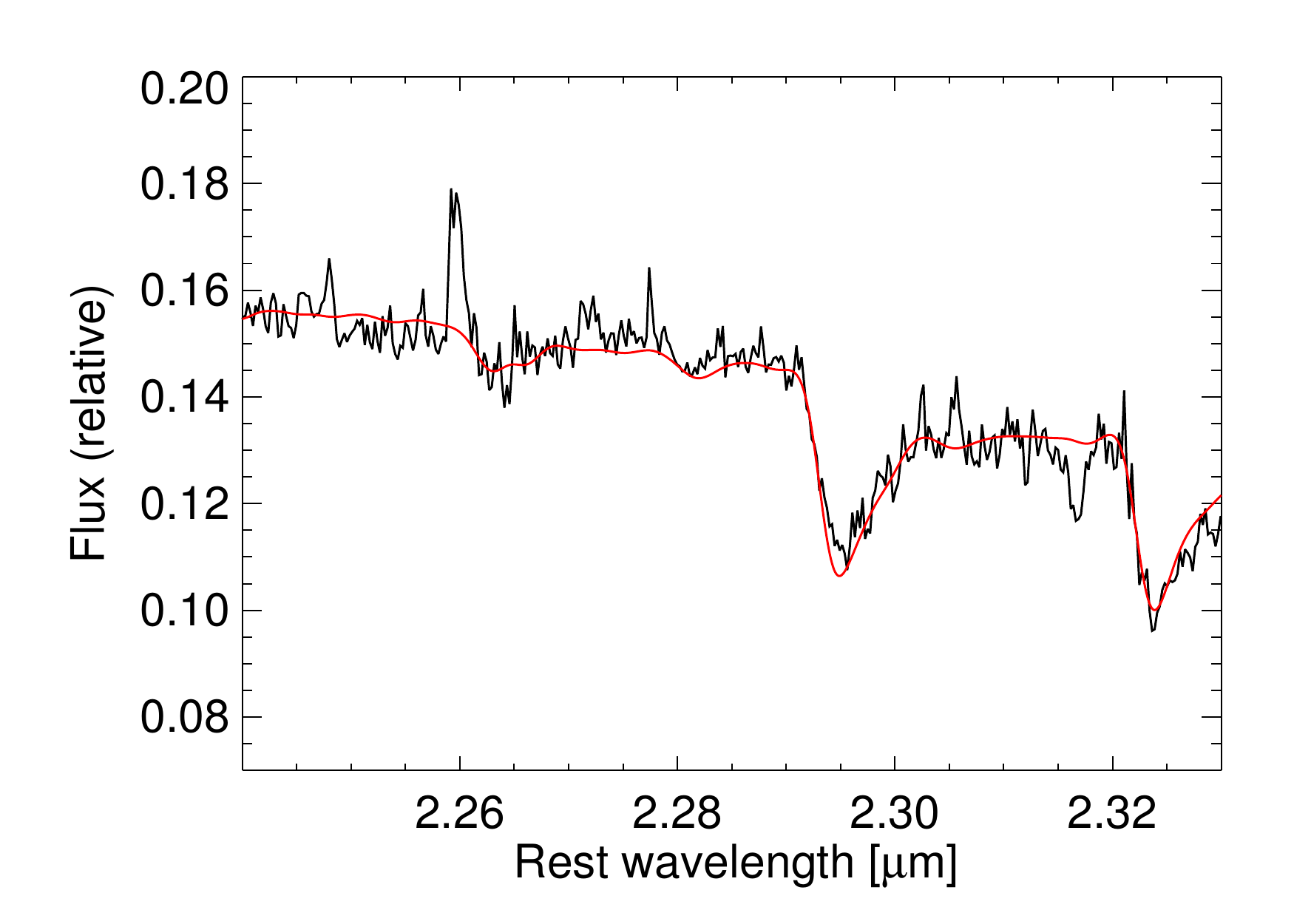}
\hspace{5mm}
\caption{\label{cofits} The black curve shows a section of the observed K-band spectrum of the Body-nuc aperture.  The red line is the best fit using a stellar template of HD109655 of type K5III.
}
\end{figure}

We fitted stellar giant and supergiant templates \footnote{http://www.gemini.edu/sciops/instruments/nearir-resources/spectral-templates} \citep{Winge2009} to the K-band 1D spectra around the CO band-head absorption region to assess the kinematics of the stellar component. Figure~\ref{cofits} shows an example.  The latest version of the Penalised Pixel-Fitting method  \citep{ppxf} was used, which simultaneously fits for both velocity broadening and radial velocity, and includes instrumental and template resolution effects.  Templates of type K5III to M0III fitted best our spectra, with M0I also resulting in satisfactory fits, though the results are not sensitive to the choice of the template within this range.  See Table~\ref{mainkinematics} for results.  For the Head, we extracted smaller apertures than the default one to avoid its rotation to play a role in the velocity dispersion, the case tabulated is from a 0.3\arcsec\ (0.3 kpc) radius circle in the middle of the Head.
The $\sigma$ measured from gas are wider than the CO stellar measurements, especially over the Body and Heart, indicating complex velocity structures and flows.

\begin{table}
  \centering
  \begin{tabular}{lcccc}
  \hline
   \hline
          &          Body-nuc  & Heart-nuc  &  Head  &  Tail   \\
 \hline
  &   & $v_r$ [km/s] & & \vspace{0.3mm}\\
 Pa$\alpha$                    &   14750$^1$  &  14470$^1$  &   14940 &  14550 \\
$H_2$ 1-0 S(1)            &    14780$^1$  &   14500$^1$  &  14920  &  14540 \\
CO band                      &   14780           &   14540         &  15000  &  14550  \\
H$\alpha$                    &   14590$^2$  &   14520$^2$  &  14940  &  14470   \\
\hline
  &  &  $\sigma$ [km/s] & & \vspace{0.3mm} \\
Pa$\alpha$    &  180 & 140 & 100 & 90 \\
1-0(S1)           & 220 & 120 & 100 & 90  \\
CO band        & 150 & 130 & 120$^3$ & 70  \\
   \hline
  &  &  $v_{rot}$ [km/s]  & & \vspace{0.3mm} \\
Pa$\alpha$    &  170 & 270 & 100 & -- \\
   \hline
  &  &  $m_{dyn}$  [$10^{10} \, M_{\odot}$] & & \vspace{0.3mm} \\
              &  10  & 7  &  2 &  -- \\
   \hline
\multicolumn{5}{l}{$^1$ Multiple velocity components, strongest one adopted. }\\ 
\multicolumn{5}{l}{$^2$ From V08. }\\
\multicolumn{5}{l}{$^3$ A small 375 kpc radius area in the Head.}\\
\end{tabular}
  \caption{\small Kinematics of the Bird per component of the system, heliocentric line-of-sight velocities $v_r$, velocity dispersions $\sigma$, and rotational velocities $v_rot$, determined along kinematic axes marked in Fig.~\ref{pa_vel}.  Uncertainties are in the range $10-20$ km/s.  Note that the Body and Heart velocity structures comprise of multiple components, increasing especially the emission line based $\sigma$ values.   Dynamical masses are calculated as in V08 from $\sigma$, $v_{rot}$ and effective radii listed therein.  
 }
\label{mainkinematics}
\end{table}

\begin{figure*}
\includegraphics[width=8cm,clip=true]{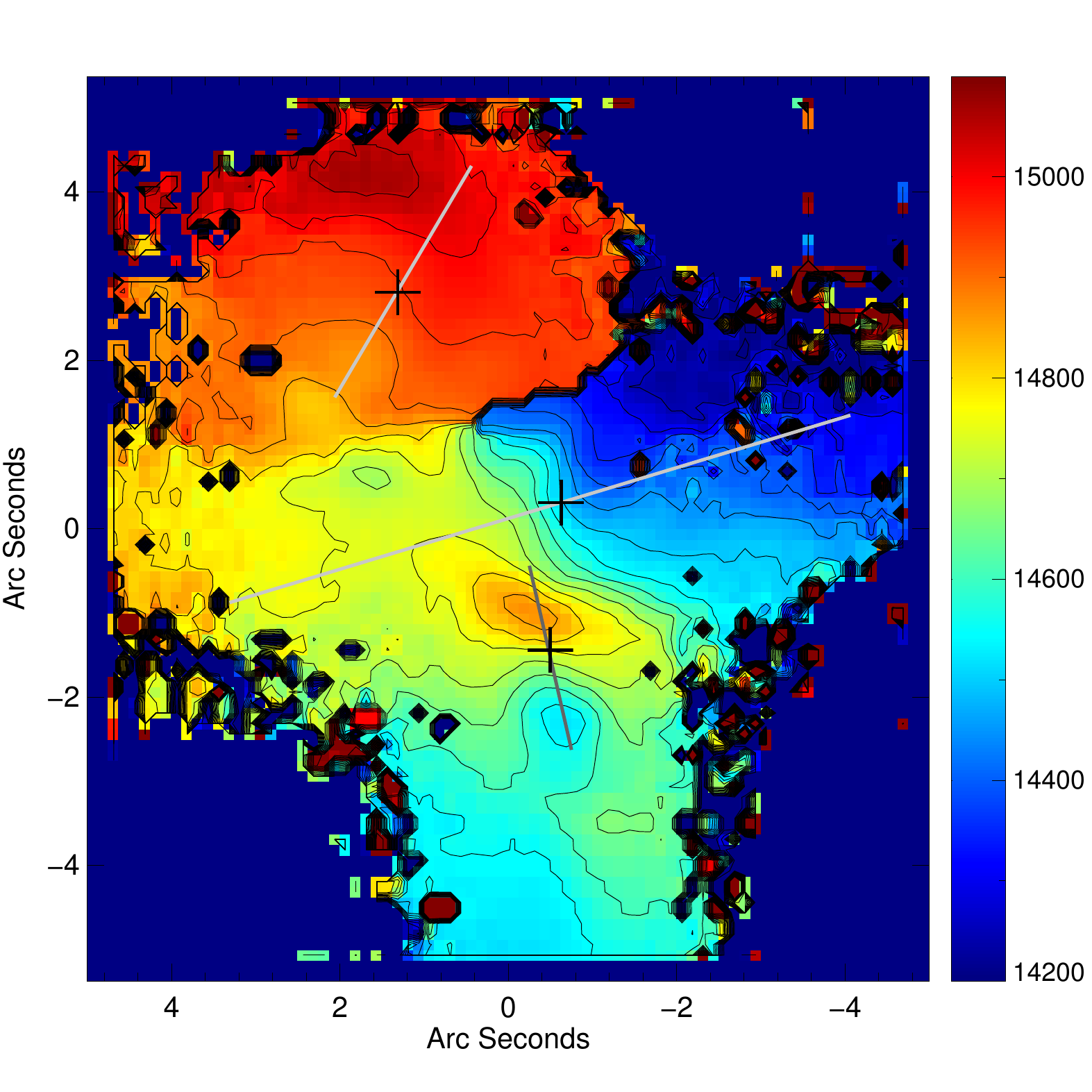}
\includegraphics[width=8cm,clip=true]{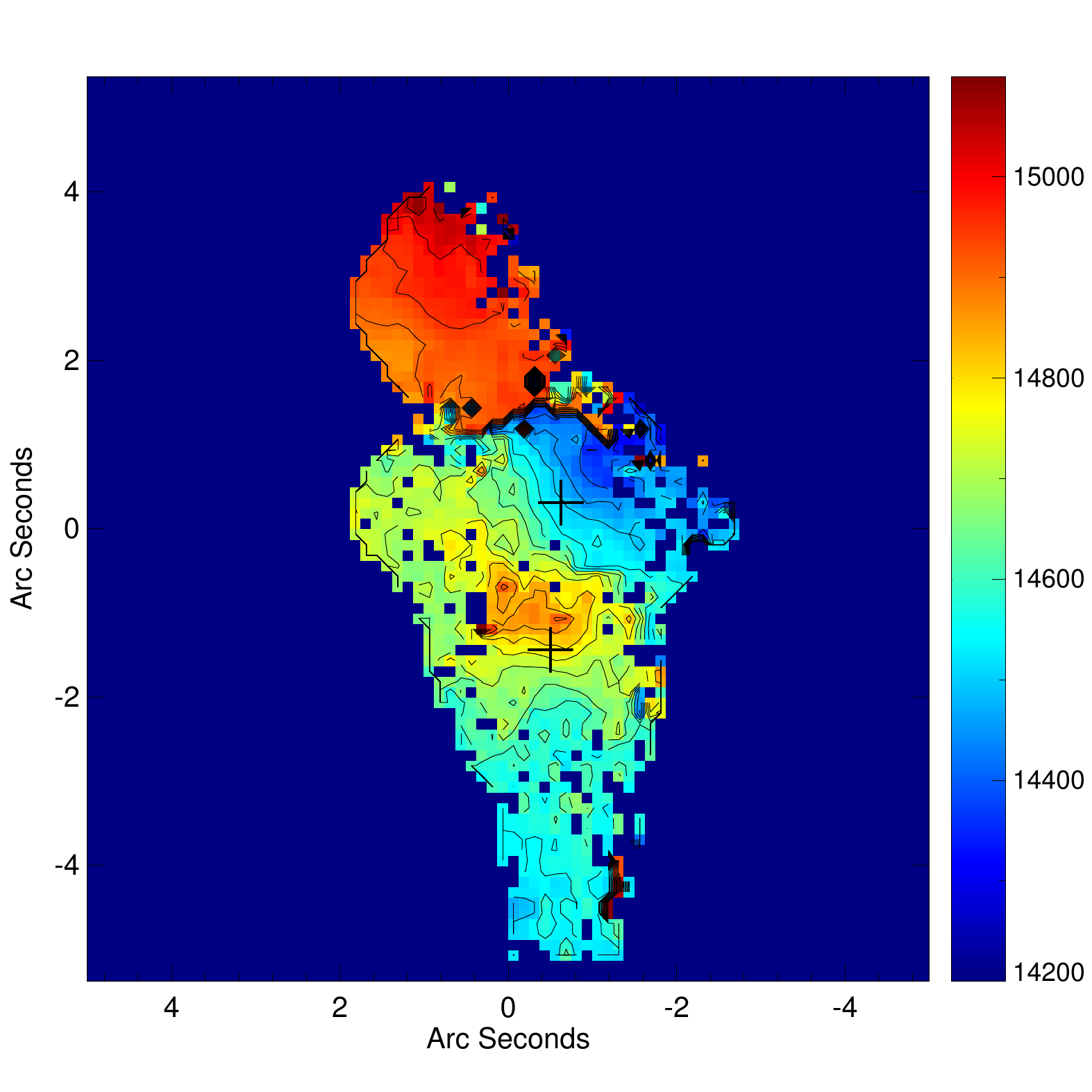}
\caption{\label{pa_vel} 
The Bird velocity field as determined from the Pa$\alpha$ line (left) and the H2 1-0 (S1) line in the K-band SINFONI observations. The Head appears to be a kinematically separate component from the rest. The crosses mark the positions of the Heart and Body nuclei in continuum light, while the cross in the Head corresponds the peak of Pa$\alpha$ emission.  The grey lines show the axes used to extract maximum velocity curves of the Head, Heart, and Body (see text).}
\end{figure*}

\begin{figure*}
\includegraphics[width=9pc]{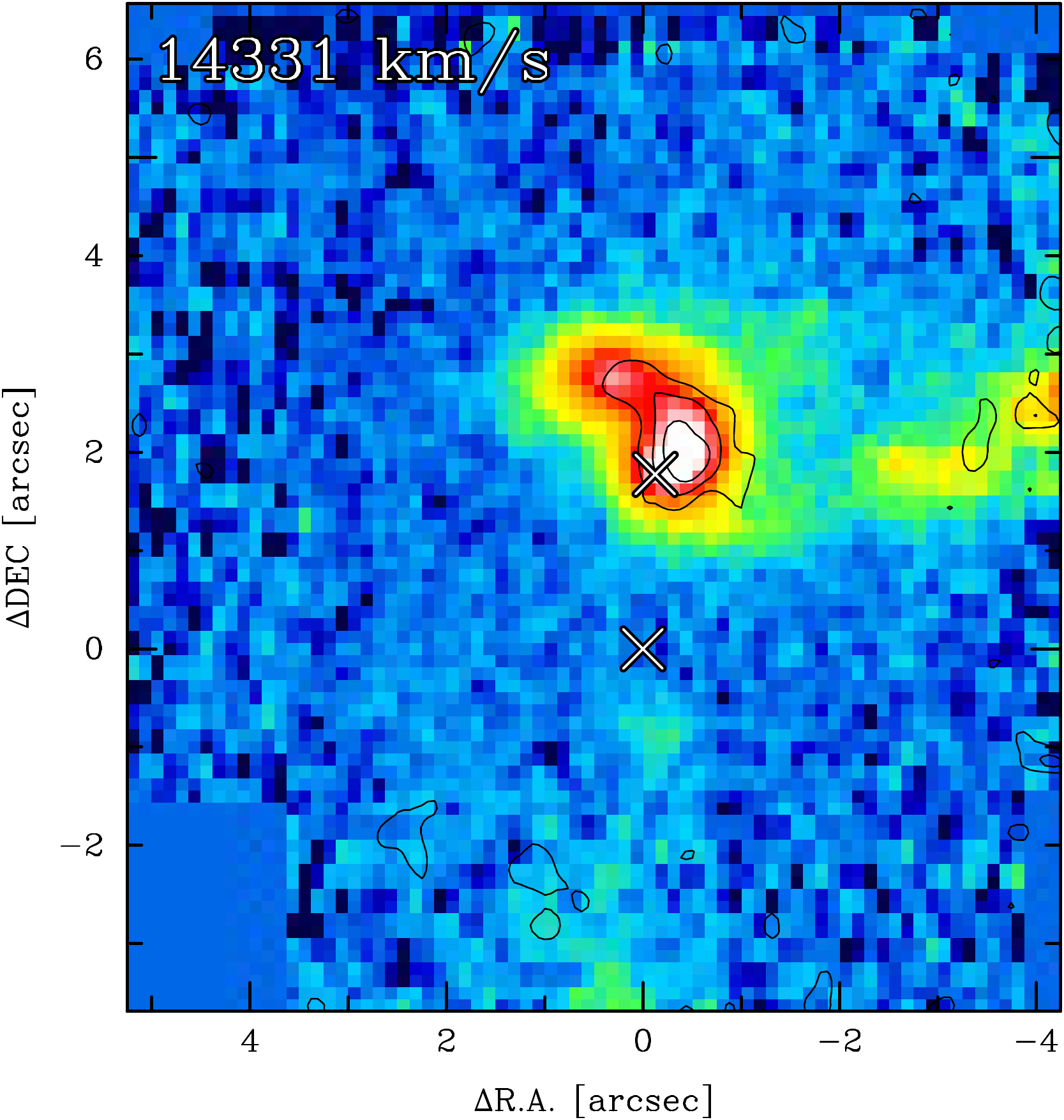}
\includegraphics[width=9pc]{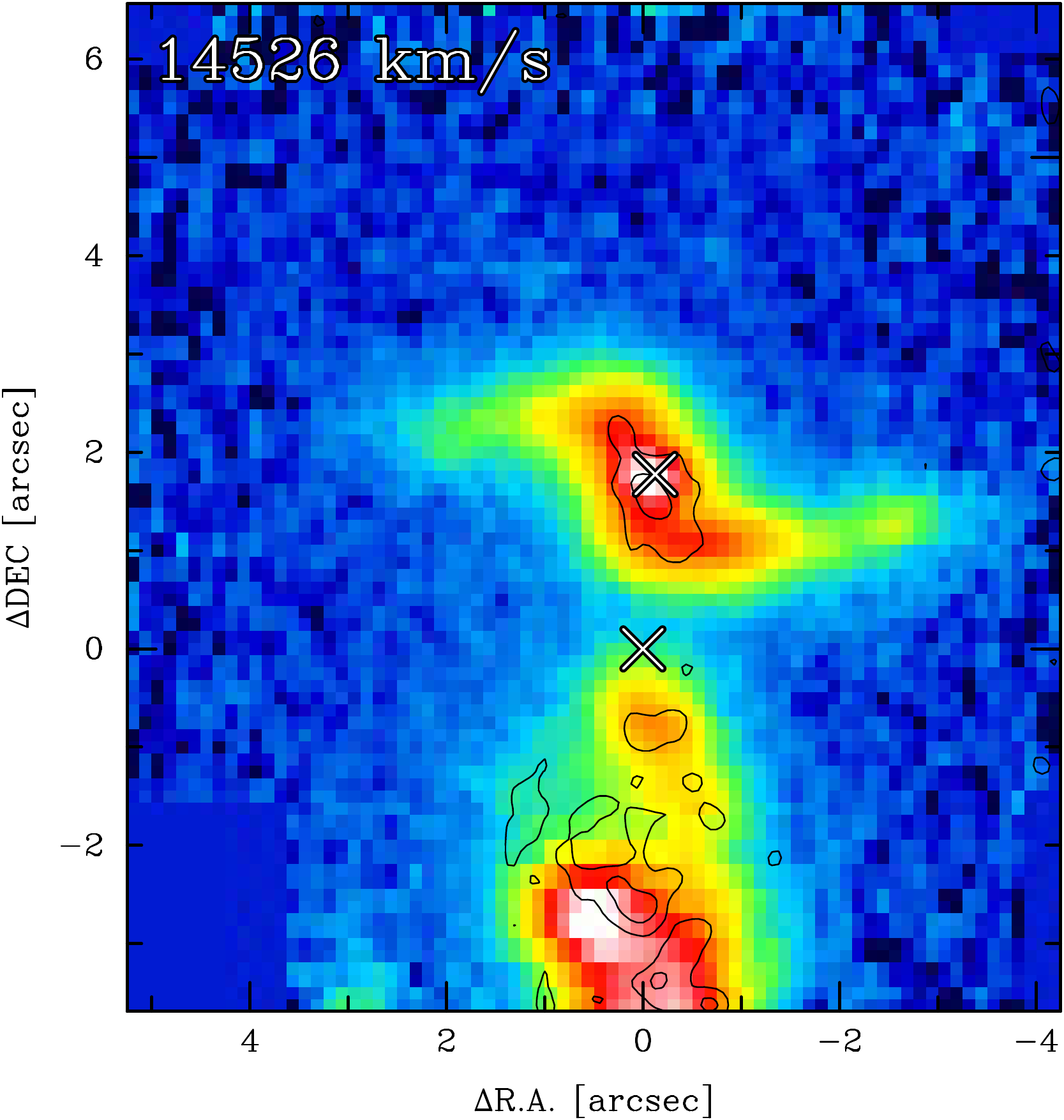}
\includegraphics[width=9pc]{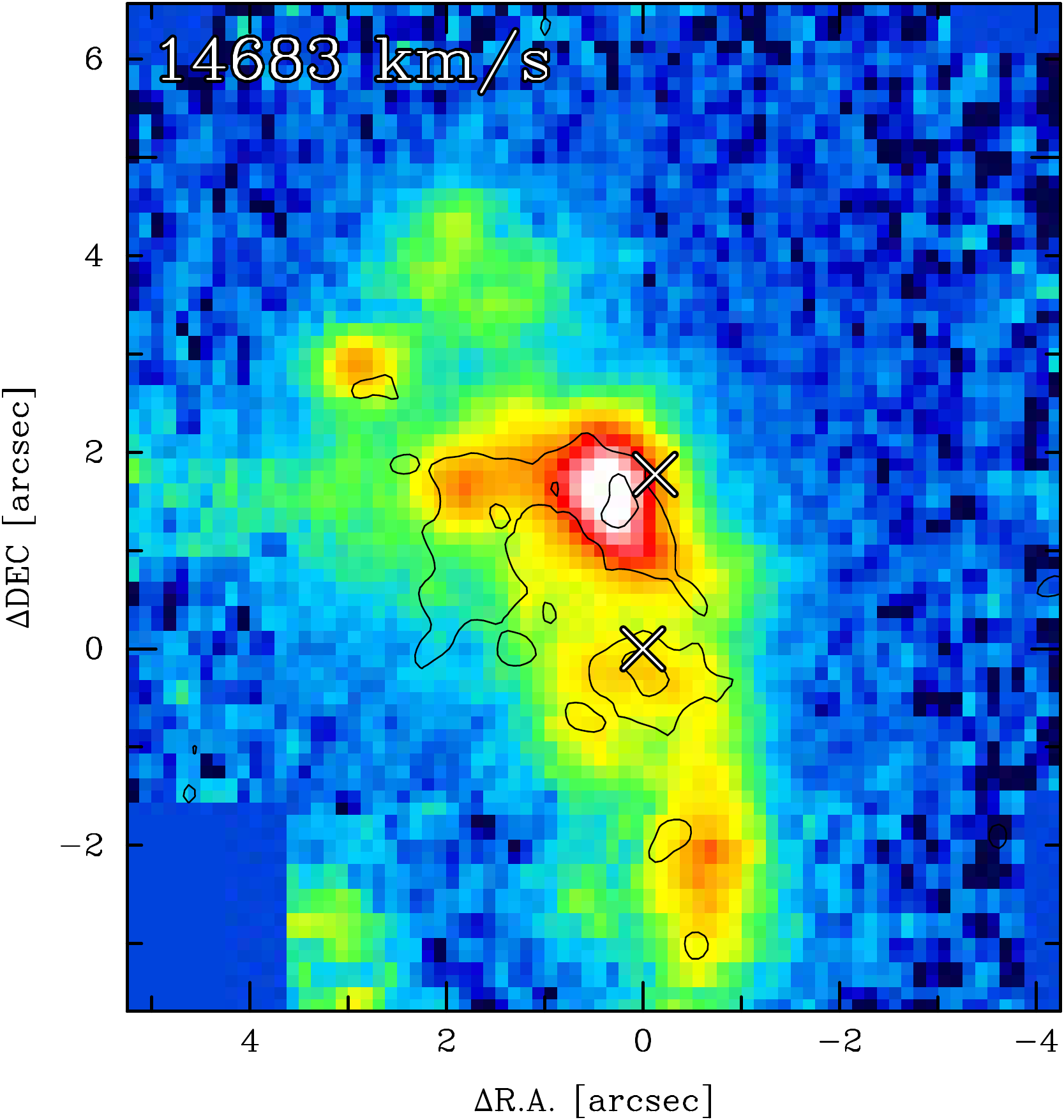}
\includegraphics[width=9pc]{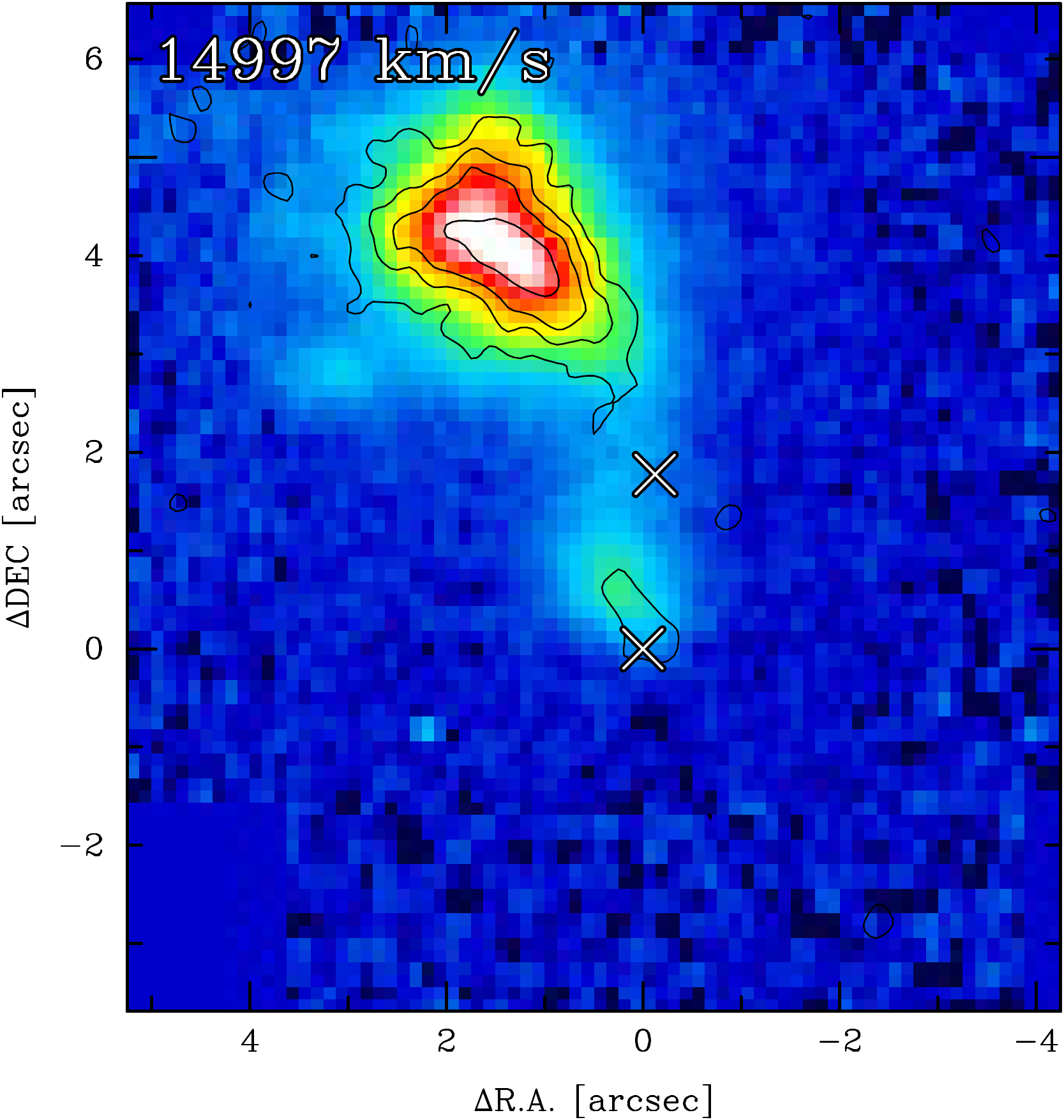} \hspace{4pc}%
\caption{\label{velfigs} \small Pa$\alpha$ maps at selected velocities from our SINFONI data, with the corresponding H$_2$  emission shown as contours.  The latter follow the Pa$\alpha$ distribution very well over the Head component, but the distributions are more varied over the main nuclei. The positions of the two main K-band nuclei seen in Fig.~5 above, are marked with crosses.  }
\end{figure*}

Secondly, full velocity maps were constructed from the IFU data cubes.  The left panel  in Fig.~\ref{pa_vel} shows the velocity map calculated by Gaussian fitting to the Pa$\alpha$ line.  Though the velocity fields are very complex especially in the central areas, the general picture follows the results of V08 based on H$\alpha$ kinematics:  the rotating Heart component connects horizontally to the tidal tails in the East (receding) and West (approaching), while the  
Head appears abruptly disjoint from the Heart, especially in the West.  However, the IFU data show that these two components have a smoother transition over the "East Wing",  which visually shows dusty material most likely connected to the Head overlapping with the tidal tails connecting to the Heart (see Fig.~\ref{bird}). 
  It is difficult to trace global North-South velocity gradients due to multiple velocity components, though the Body and Tail do clearly belong to the same structure based on imaging and long-slit spectra (Figs.~\ref{bird} and~\ref{saltspec}), and apparent disk rotation is detected in the Body location (see below).    A velocity field constructed from the H$_2$ 1-0 S(1) line, in the right panel of Fig.~\ref{pa_vel}, shows the same trends as the Pa$\alpha$ version.   The general complexity of the kinematic field over the Bird galaxy is illustrated in the Fig.~\ref{velfigs} where we show the distribution of both ionised and molecular gas at a few selected velocity channels. The velocity range of both ionised and molecular gas over the Heart and Body nuclear locations of the Bird is very large.  The decomposed velocity structures in different spatial positions, relating especially to gas flows in the system, are analysed further in the separate follow-up work in V\"ais\"anen et al. (in prep.).  

An important feature not detected in the V08 long-slit spectroscopy, but obvious in the maps presented in Fig.~\ref{pa_vel}, is the high-velocity area in the middle of the field-of-view in between the Body and Heart components, and a corresponding low velocity area just south of the Body nucleus.  The simplest explanation of it is a fairly high inclination rotating disk, embedded in the other velocity fields.  It was not detectable in the optical SALT data at all due to obscuration, but spatially it matches well the elongated shape of the K-band AO-image of the Body and the "bar-like" structure left over when a bulge+disk surface brightness profile was subtracted off it (see fig.\ 6.\ in V08).  The radius of this rapidly rotating inner disk is $\approx 1.0$ kpc, and rotational velocity $v_{rot} \approx 170$ km/s without inclinations corrections.  
The approximate kinematic PA is marked in Fig.~\ref{pa_vel}, as derived from the Kinemetry IDL program\footnote{http://davor.krajnovic.org/idl/} of \citet{kinemetry2006}.  The approximate kinematic PAs of the Heart and the Head are also marked, and the $v_{rot}$ values determined as half of the maximum velocity amplitude along the relevant axes are 270 and 100 km/s, respectively.  No further fitting or modelling of the velocity maps is attempted in this work, however, due to the complexity of the system.  As discussed in Section~\ref{kinedisc}, all things considered, we argue that a three-galaxy interaction makes most sense, even though a very distorted two-galaxy scenario cannot be ruled out without detailed modelling.   

The stellar velocity dispersion $\sigma$ measurements, and rotation, allow an estimate of the dynamical masses (Table~\ref{mainkinematics}), to compare with the values derived in V08 based on ionised, optical, gas dispersion and rotation measurements.  The Body $\sigma =140$ km/s together with the rotation and an effective radius of $r_e \approx 2.4$ kpc indicates a mass of $\sim 1 \times10^{11} M_{\odot}$, somewhat larger than our old estimate. The Heart is more strongly dominated by rotation.  Using the CO-based $\sigma$ and rotation (Table~\ref{mainkinematics}), and a $r_e \approx 1.3$ kpc, the dynamical mass remains the same than in V08, $\sim 7 \times10^{10} M_{\odot}$.  The Head is the most difficult to assess, though easier now with the IFU data. Its shape is not well-behaved, but the photometric half-light radius is calculated to be $r_e \sim 0.8$ kpc.  It would not be appropriate to measure the $\sigma$ in the default large Head aperture due to clear rotation (Fig.~\ref{pa_vel}) inside it.  Rather, we fitted the CO-line region from multiple smaller apertures within the Head:  an interesting feature is that we derive a surprisingly high $\sigma \approx 120$ km/s in a small area in the middle of the Head, while elsewhere the velocity dispersion tends to be in the range 70-90 km/s. The rotation of the Head is more defined, and larger, in fact, than determined in V08, with approximately $v_{rot} \sim $100 km/s estimated from a velocity curve extracted with PA $\approx -20$ marked by the line across the Head in Fig.~\ref{pa_vel}.  These values double its dynamical mass estimate to $\sim 2\times10^{10} M_{\odot}$.

Since dynamical masses are essentially calculated within half-mass radii, crude {\em total} stellar masses can be estimated as double those derived above and listed in Table~\ref{mainkinematics}.  The main progenitors of the Bird system clearly are very massive $1-2 \times 10^{11} M_{\odot}$ disk galaxies.  The implied dark matter halo sizes are thus likely slightly in excess of  $10^{13} M_{\odot}$  \citep{Moster2013}.

\subsection{Stellar populations and metallicities}
\label{stellarpops}

\begin{figure}
\includegraphics[width=8cm,clip=true]{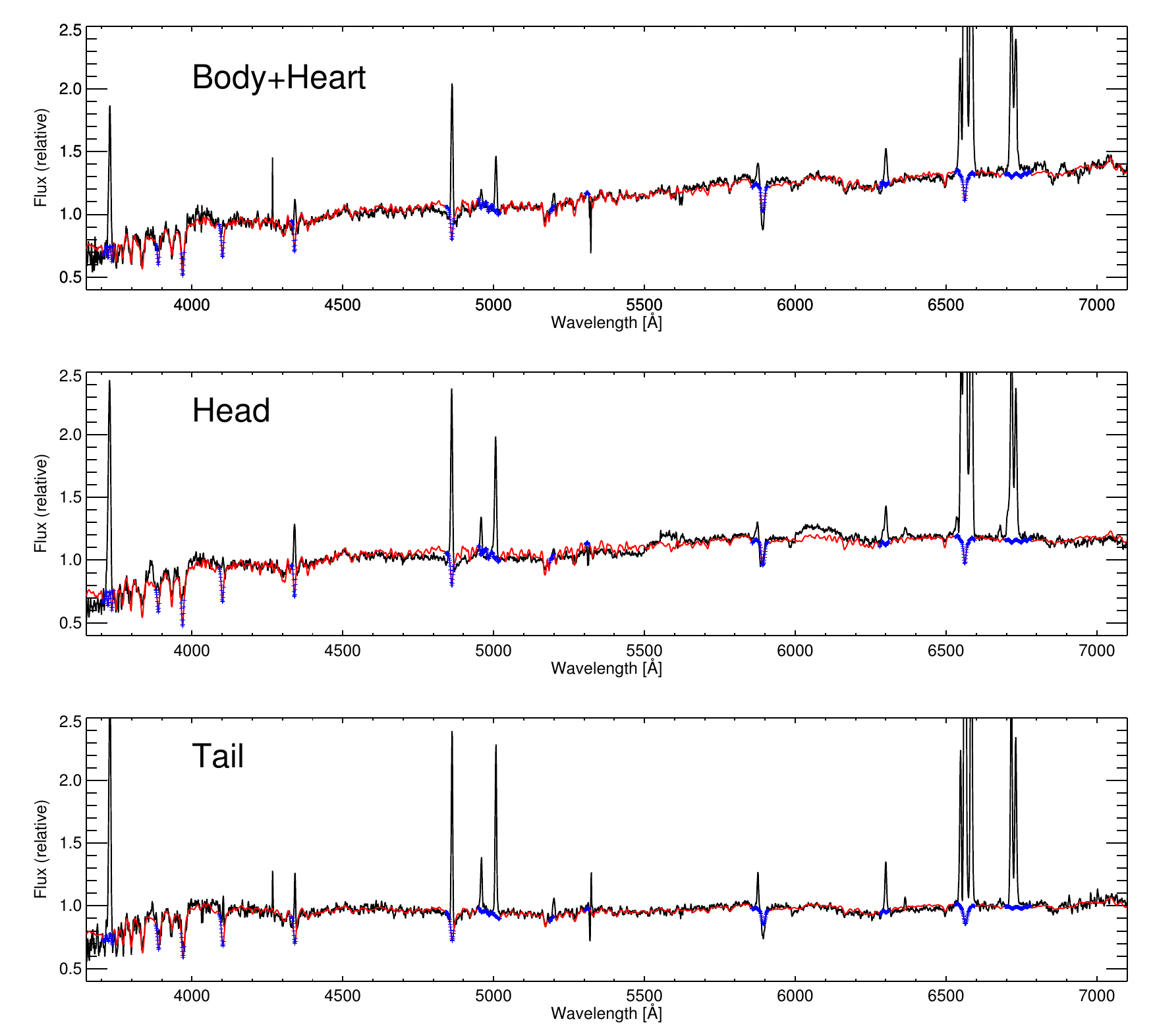}
\hspace{5mm}
\caption{\label{starlight}  The extracted SALT spectra of the Body+Heart, Head, and Tail, are shown as the black curves, the red curves depicting the best-fit SSP model using Starlight  (see text).  The small blue crosses show pixels which have been masked out of the fit due to emission line contamination.
}
\end{figure}

\begin{figure}
\includegraphics[width=8cm,clip=true]{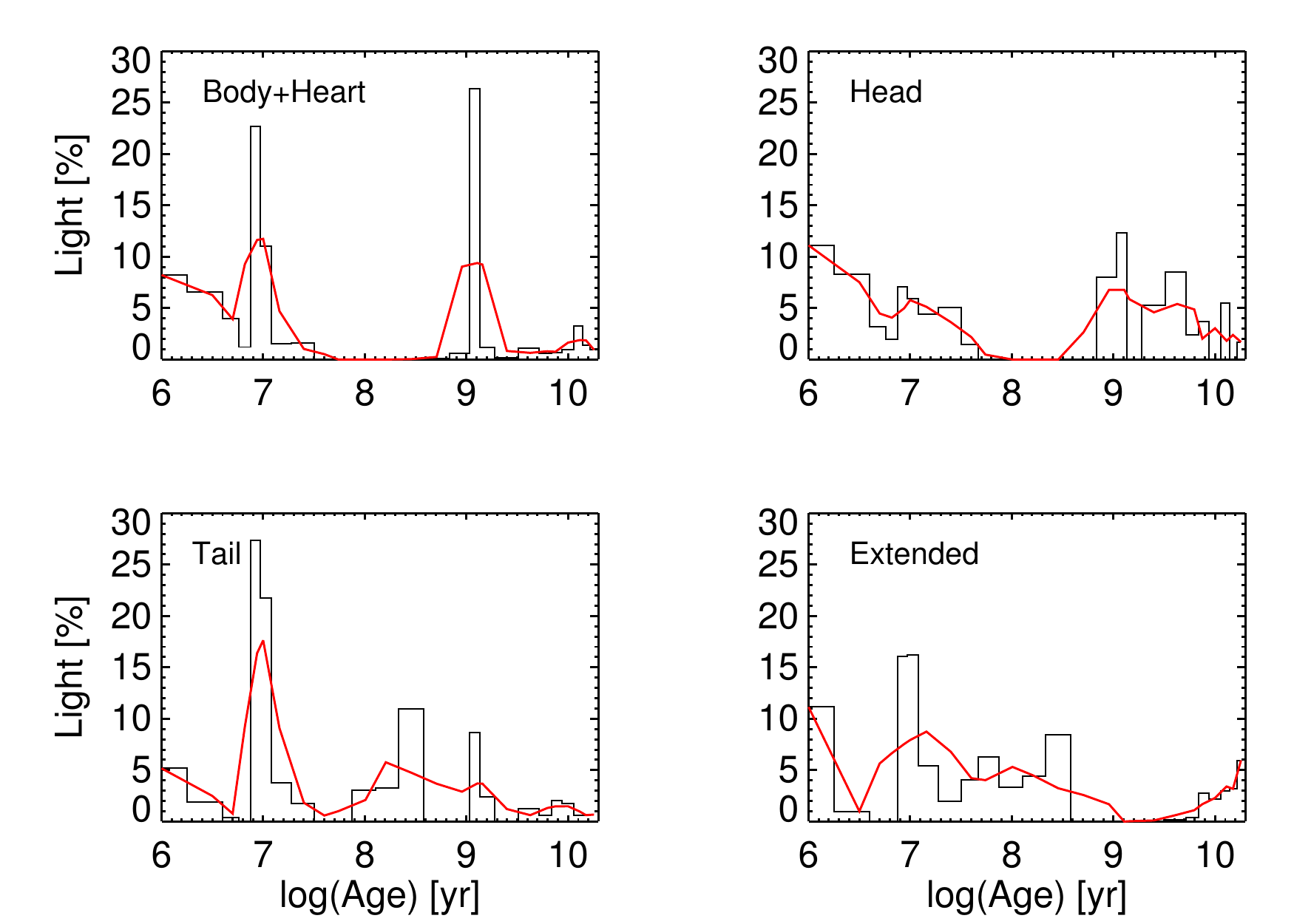}
\hspace{5mm}
\caption{\label{sfh}  The light fractions originating from SSPs per age in the best-fit Starlight runs per Bird component.  The plots are not intended to be accurate star formation histories, but rather indicative of the evolution.   The red curves are 3-bin smoothed versions of the histogram.
}
\end{figure}

The new optical SALT spectrum was fit using the Starlight package \citep{CidFernandes2005} in conjunction with BC03 SSP models \citep{BC03} 
using the Padova tracks. The fit was done separately for apertures corresponding to the Head, 
Body+Heart (a mixture of the components which could not be well separated due to the spatial resolution of the data), 
Tail, 
and the low surface brightness emission region north of the Head (see Fig.~\ref{saltspec}). 
Figure~\ref{starlight} shows the extracted Tail, Head and Body+Heart spectra, together with their Starlight fits. The fit results are tabulated in Table~\ref{slfits}. 
All the light-weighted mean ages between the apertures are in the range between 0.9 and 2.4 Gyr, though there is strong contribution from very young ages making average values less meaningful.  An old underlying population is present everywhere, as expected, and it also dominates the mass -- the mass-weighted mean age is close to 10 Gyr for all components.  Figure~\ref{sfh} plots the SSP fractions that Starlight picks out for the best-fit model, as an indicative star formation history.     
The fractions of light coming from various bins of ages are also tabulated in Table~\ref{slfits}.   As is seen, approximately half of light originates from a very young stellar population of $<15$ Myr in all the apertures, and the recent SF appears similar throughout the Bird.

\begin{table*}
\begin{footnotesize}
\vspace{0.5cm}
\begin{tabular}{llccccccc}
\hline
\hline
   &  age$_{light}$ $^a$ & $Z$  & A$_V$  (stars) &  $<15$Myr &   $15-800$Myr  &  $800-1400$Myr  &  O/H  &  A$_V$ (gas) \\
   &  (Myr) &   &  (mag)  &  (\% light)  &  (\% light)  &  (\% light)  &  &  (mag)  \\
\hline
Body+Heart &  1300 / 60    &  0.016  &  1.6   &  55 &  2  &  27  &  8.78  &  3.0 \\
Head            &  2300 / 146  &  0.011  &  1.1   &  42 &  7  &  20  &  8.79  &  2.5 \\
Tail               &  900 / 44      &  0.019  &  1.5   &  60  & 19  &  9  &  8.74  &  2.4 \\
Extended     &  2400 / 45    &  0.018  &  1.2   &  50 &  29  &  0  &  8.85  &  1.2\\
\hline
\multicolumn{6}{l}{$^a$ First value is the average in linear space, the second in log space.}
\end{tabular}
\caption{\label{slfits} Selected stellar population and ionised gas characteristics in the Bird derived from the SALT long-slit spectra using BC03 models for delta-function star bursts.
The {\em light weighted} mean age is not very meaningful in case of more than one widely separated discrete star-formation events, which is reflected in differences whether the mean is taken in linear or log space.  Mass weighted ages (not tabulated) are all close to 10 Gyr.   
Columns 3 to 4 show the best-fit average stellar metallicity and extinction, and columns 5 to 7 the 
percentages  of light coming from stellar populations in the indicated age ranges.  Columns 8 and 9 show the Oxygen abundance and extinction of ionised gas derived from the strong emission lines of the spectra.}
\end{footnotesize}
\end{table*}

Are there any differences between the apertures?  The spectra in Fig.~\ref{starlight} are broadly similar, though there are differences in the details of the Balmer jump region, as well as the Mgb and Fe line regions. 
Looking at the resulting SFH plots in Fig.~\ref{sfh}, the Body+Heart SFH plot shows a significant peak of SF around 1 Gyr, while the other apertures have somewhat more extended distribution of SF.   The Tail and the extended emission North of the Head have more intermediate 100 Myr to 1 Gyr old star formation, while the most massive nuclei appear quite devoid of SF in this time scale. Note that it is not possible to fit the population of the Body by itself, both because of spatial resolution, but also because of its obscuration.  The differences in age distributions remain if we use a fixed metallicity (e.g.\ Solar) in the Starlight fits, or let it vary as is the case in the plots shown.  These results are discussed in Section~\ref{histories}. 

It is especially interesting that the stellar [Fe/H] metallicity of the Head appears lower than in the other apertures, by approximately 0.2 dex. This difference is, incidentally, similar to what is expected from a galaxy with a factor of 3 to 5 lower mass than the Heart or the Body \citep[see e.g.][]{Gallazzi2005}.

The strong emission line ratios, after subtracting the best-fit models from the spectra, were also used to derive nebular abundances using common empirical methods. The gas-phase metallicities show uniformly (Table~\ref{slfits}) slightly over Solar values over all the apertures with approximately $12+\log ( [O / H]) \approx 8.8 \pm 0.1$ using averages of O3N2, DO2, and N2 methods adopted from the formulations of \citet{Kewley2008}.  These values are fully consistent with the abundance derivation from optical IFU spectroscopy (WiFeS) presented by \citet{Rich2012}, which, while shallower in S/N, also probe the wings of the Bird that our long-slit data do not cover.  The measured abundance is just below values expected for galaxy masses derived above, depending on which mass-metallicity relation calibration is exactly adopted \citep[e.g][]{Kewley2008}.  The line-ratios are typical to HII regions, except in the the large ionised gas structure extending nearly 30\arcsec, or 30 kpc, North from the top of the Head (see lower panel of Fig.~\ref{saltspec}): the [NII]/H$\alpha \sim 0.6$ and [SII]/H$\alpha \sim 0.6$ ratios are elevated here, suggesting more dominant shock ionisation in the extended emission.

The A$_V$ extinction was derived together with the Starlight fits for the stellar population, and as a by-product of the iterative Oxygen abundance determination for the gas-phase extinction.  The best-fit stellar population A$_V$ values are between 1.1 and 1.6 magnitudes, while the gas-phase values are 2.4 to 3.0 mag, except a low of 1.1 for the emission gas north of the Head component.  Such differences are typical in star-forming regions.  All these A$_V$'s are significantly lower than the NIR derived values in Table~\ref{lineratios}, however, which is not surprising since the optical observations do not probe as deep into the dusty LIRG as NIR does.

There are several NIR absorption lines detectable in the J and K data-cubes, apart from the CO bands already discussed, to be used for further constraints on the stellar populations.  Ignoring the Na I lines at 1.1395 and 2.2063 $\mu$m due to possible ISM contamination, these are: Al I at 1.1250 $\mu$m, Si I at 1.2112    $\mu$m, and Ca I at 2.2655 $\mu$m.  We measured the equivalent widths  of these features (Table~\ref{eqwid} which also shows emission line EWs discussed later in Section~{sbage}) in each of the  apertures; determining the continuum is not trivial making the values approximate. The Body and Heart apertures show similar ranges with EW(Al I) $\sim 1-2$ \AA, EW(Si I) $\sim 2-4$, EW(Ca I) $\sim 1-2$, and EW(CO 2.3$\mu$m) $\approx 14$.  Comparing the numbers to models of \citet{Maraston2005} as discussed in \cite{Riffel2008}, the values clearly indicate a dominant $\sim$1 Gyr population. The Tail has weaker absorption features of the three metal lines, but still detectable with EW $\sim 1-2$.  In contrast, the Al I, Ca I and Si I features in the Head aperture are only marginally detectable at EW $< 0.7$ \AA\ level, or not at all, while the CO is nearly as strong as elsewhere with EW(CO 2.3$\mu$m) $\approx 11$:  this situation is satisfied with a dominant young $\sim10$ Myr population. These results are consistent with the view that the Head has a different history compared to the other components.  It is notable that practically no absorption lines are present in a spectrum extracted from the location of the peak of the Head recombination emission flux, suggesting extremely young SF therein.

\begin{table*}
\label{eqwid}
\begin{footnotesize}
\vspace{0.5cm}
\begin{tabular}{lcc|cccccc}
\hline
\hline
   &  CO 2.30 $\mu$m  &   Al I 1.13 $\mu$m & Si I 1.21 $\mu$m  &  Ca I 2.27 $\mu$m &  Pa$\alpha$  &  Br$\gamma$  &  SB age [Myr] & SP age [Myr] \\
\hline
Body-nuc         &  19      &  1.2        &  2      &  1.5     &  14     &  $<2$  &   40    &   1000  \\
Heart-nuc         &  18       &  1.2       &   3     &   2.5     &  85    &  13      &  8       &  1000  \\
Head               &  13       &  $<0.6$ &  --      &  $<0.5$ &  220  &  31      &   6--7 &   10  \\
Head-peak      &  $<7$   &  --         & --        &  --         &  720  &  105      &   4      &    $<9$  \\
Tail                  &  16       &  1.3       &  $<1$  &  $<1$   &  95    &  14      &   8     &   10, 1000  \\
\hline
\end{tabular}
\caption{Equivalent widths, all in \AA, tabulated in selected apertures from the SINFONI J and K datacube.  The Head-peak refers to the location of peak Pa$\alpha$ flux within the Head.  First four columns are absorption lines.  The "SP age" refers to a dominant age of an underlying stellar population based on the absorption line EW's (see text, Section~\ref{stellarpops} and the "SB age" to a dominant age of an instantaneous starburst based on EW(Br$\gamma$) and CO index (see text, Section~\ref{sbage}). }
\end{footnotesize}
\end{table*}

\section{Discussion}
\label{discussion}

\begin{figure*}
\includegraphics[width=7cm,clip=true]{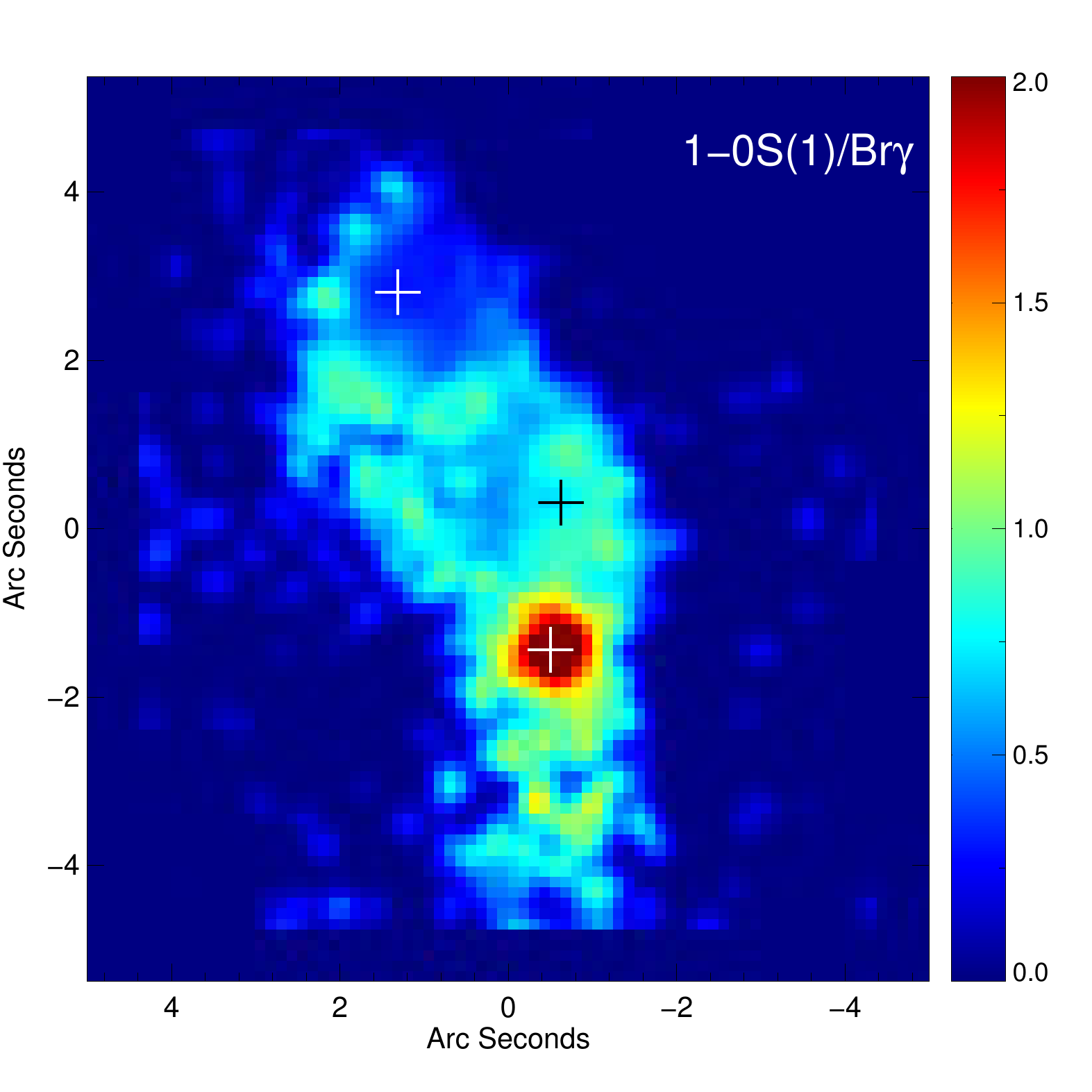}
\hspace{2mm}
\includegraphics[width=7cm,clip=true]{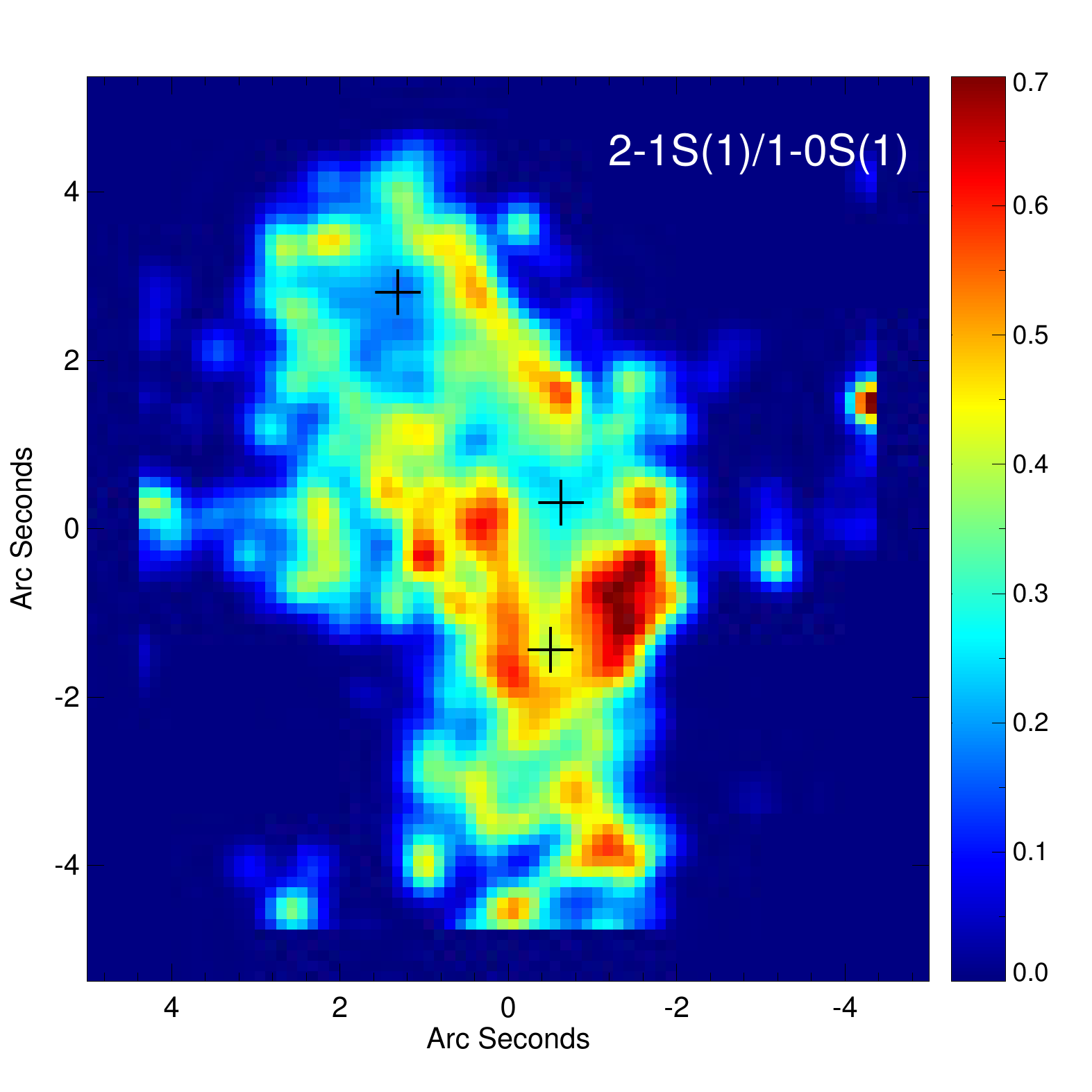}
\hspace{2mm}
\includegraphics[width=7cm,clip=true]{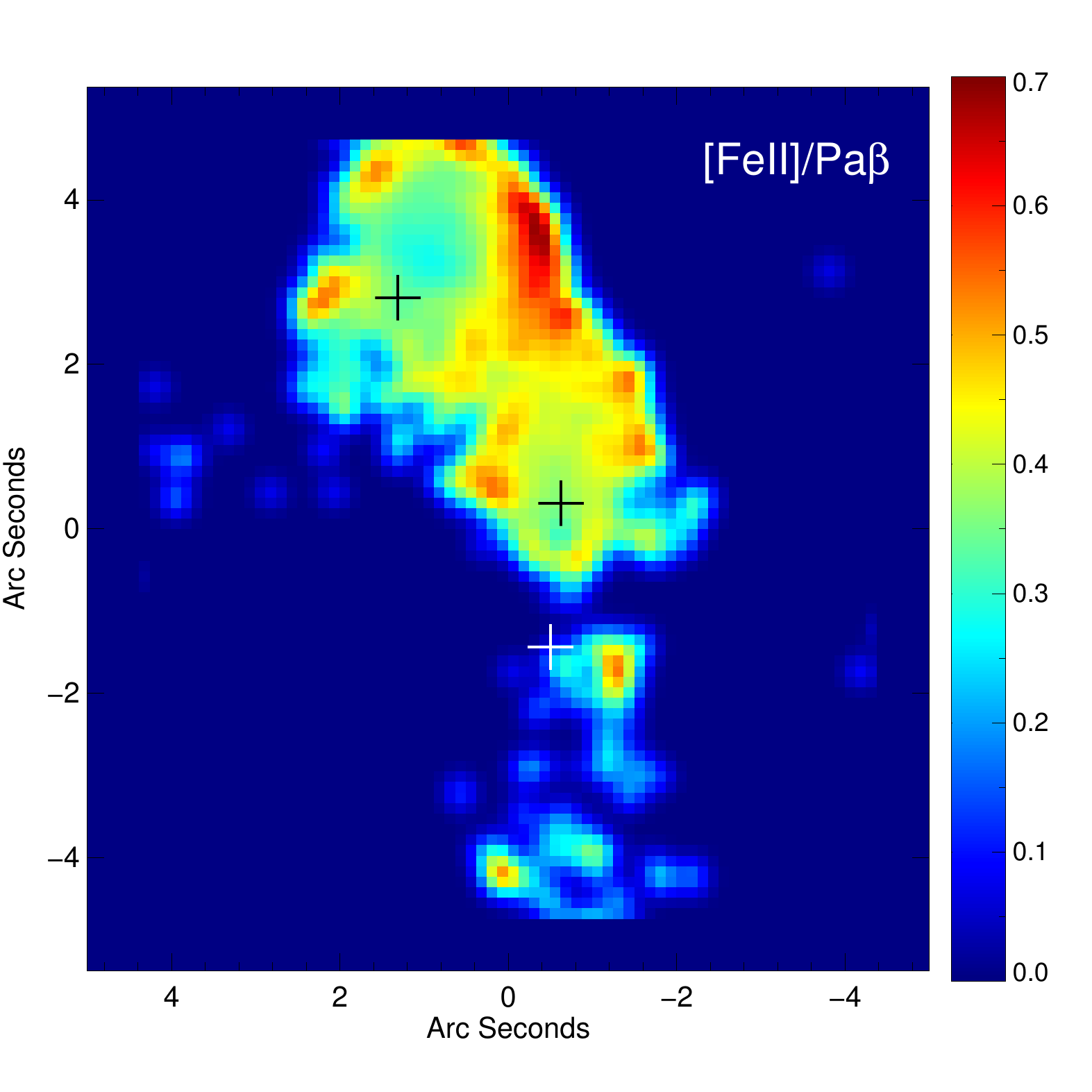}
\hspace{2mm}
\includegraphics[width=7cm,clip=true]{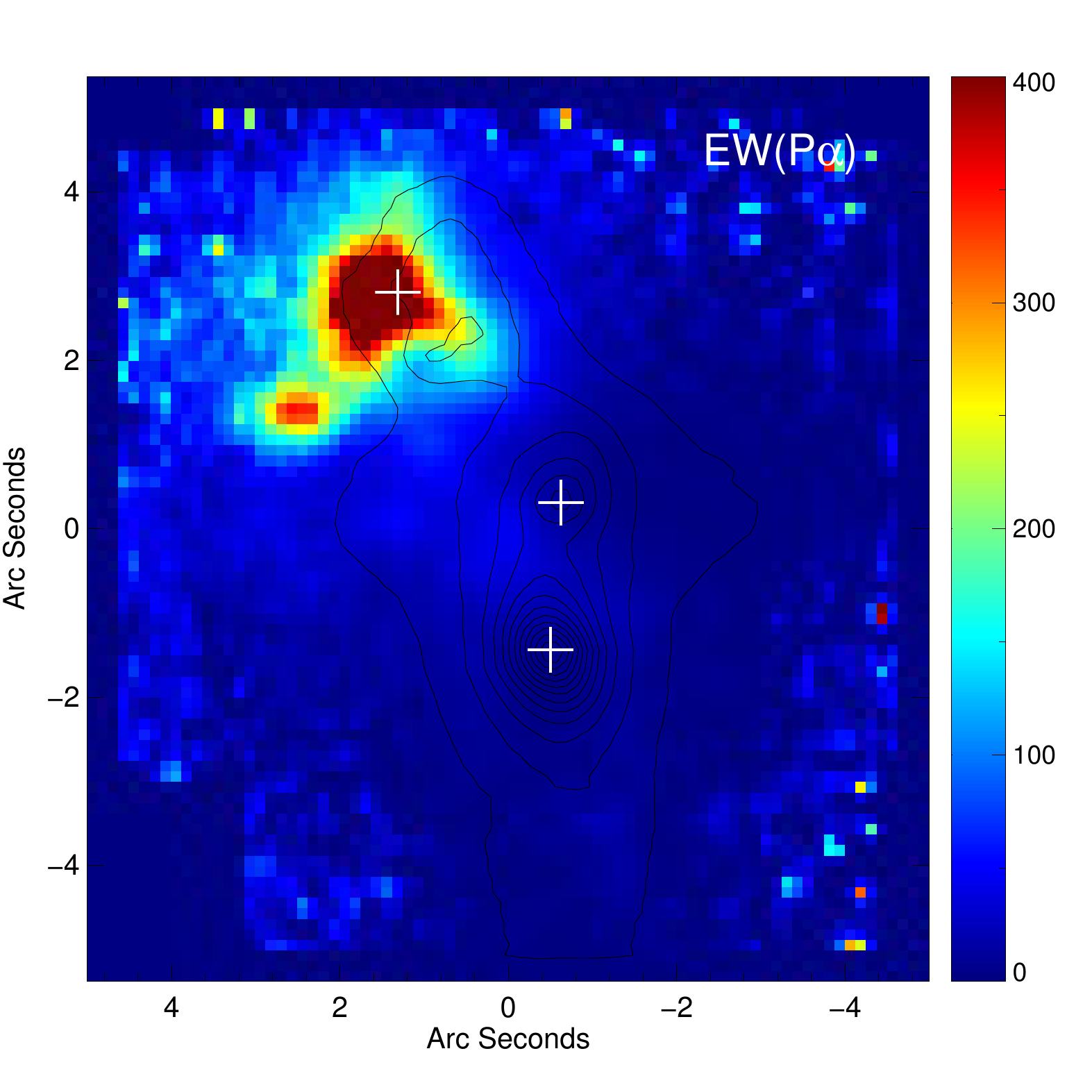}
\caption{\label{ratiomaps} Selected line-ratio maps,  smoothed with a 2.5 pixel width Gaussian, and an equivalent width map.  The [Fe II] / Pa$\beta$ is from the SINFONI J250 data-cube, the others from the K250 data. In the H$_2$ 2-1 (S1) / Br$\gamma$ ratio map, the Body peaks at values of approximately 4, while in the Pa$\alpha$ EW map the peak Head locations has values in excess of 700 \AA.  
}
\end{figure*}

A puzzling feature of the Bird galaxy is the lack of very strong star formation in the main nuclei as expected from the morphological signs of typical (U)LIRGs, and the fact that there is plenty of molecular gas throughout the system, both warm gas as shown by the $H_2$ emission we have detected, and cold gas \citep[][Romero-Canizales, in prep.]{Mirabel1990}.   The outstanding question which could not be answered previously is whether a starburst has not really even started in the most massive Body nucleus, or whether it has been quenched, and if so, why.  What are the differences between the Bird components, what are their evolutionary stages?  It is these questions that we focus on with the analysis of the observational results.  Figure.~\ref{ratiomaps} shows selected line ratio maps used in the discussion below.

\subsection{Excitation mechanisms} 
\label{exmech}

The H$_2$ line ratios measured over the Bird system (Table~\ref{lineratios}) can be used to determine the sources of excitation of the molecular gas \citep[e.g.][]{Mouri1994,Emonts2014}.   Comparisons of 1-0S(2) / 1-0S(0)  or 1-0S(3) / 1-0S(1) against 2-1S(1) / 1-0S(1) can differentiate between thermal and non-thermal processes, and also reveal excitation temperatures, which in turn reveal thermal heating mechanisms.  
We calculate temperatures using H$_2$ line ratios following equations adopted from \cite{Reunanen2002} and tabulate them in Table~\ref{temperatures}.  

The vibrational excitation temperature is calculated using extinction corrected (adopting the Br$\gamma$ / Pa$\alpha$ derived value) line fluxes:
\begin{equation} 
T_{vib} \sim 5600 / \ln(1.355 \times I_{1-0S(1)} / I_{2-1S(1)}), 
\label{tempvib}
\end{equation}
and the rotational excitation temperature from:
\begin{equation} 
T_{rot} \sim -1113 / \ln(0.323 \times I_{1-0S(2)} / I_{1-0S(0)}). 
\label{temprot}
\end{equation} 
We also tabulate the ortho/para ratio of the excited molecular gas using the S(3), S(2), and S(1) transition ratios using Equation (1) in \citet{Reunanen2007}.

The Head and Tail locations have $T_{rot} \sim 1000$ , whilte $T_{rot} \sim 2000$ K at the Heart and Body nuclei.  The vibrational temperatures are systematically higher everywhere $T_{vib} \sim 3000-4000$ K.  \citet{Riffel2013} find similar trends in their starburst LIRGs. 
The differences between $T_{rot}$ and $T_{vib}$, as well as the line-ratios involving $\nu=1$ and $\nu=2$ transitions, indicate \citep[see e.g.][]{Mouri1994} that there are non-thermal processes (UV fluorescence) involved in the excitation. The Body aperture with 2-1S(1) / 1-0S(1) $\sim 0.5$, in particular, appears to be dominated by non-thermal emission.  
 
The 1-0S(2) / 1-0S(0)  and 1-0S(3) / 1-0S(1) ratios, and the derived $T_{rot}$ values, in turn, show the dominant mechanism of the thermal component of excitation \citep[][]{Mouri1994}: the low $T_{rot} \sim 1000$ K temperatures of the Head and Tail suggest UV photon heating, such as in photodissociation regions, while the higher $T_{rot} \sim 2000$ K temperatures at the Heart and Body strongly suggest shock heating. 

Digging deeper into the data, we also measured the same H$_2$ line ratios in the same apertures, from line components after de-blending the velocity profiles.  Over the small Body aperture it is clear that the largest  2-1 S(1) / 1-0S(1) and 1-0S(2) / 1-0S(0) ratios come from the {\em blue-shifted} line emission, while the ratios at the systemic Body velocity component at 14770 km/s are similar or slightly lower than the ratios and temperatures discussed above.  The blue-shifted component ratios suggest that non-thermal UV fluorescence is actually the dominant mechanism in these outflows, while the thermal excitation temperatures are even higher than above, consistent with shock heating.  Figure~\ref{ratiomaps}, top right panel, showed that the highest  2-1S(1) / 1-0S(1) values are {\em around} the Body nucleus nearly perpendicular to the rotating disk,  strongly suggesting the ratios to be related to conical outflows from the Body nucleus. The Head and Heart have lower values closer to pure thermal emission. 
Measurements such as these pin-point why global velocity dispersions in LIRGs correlate with the ionisation level \citep{MonrealIbero2010}, as do the $\sigma$ values of the main apertures in our case (Table~\ref{mainkinematics}) --- large $\sigma$ indicate outflows which are shock heated.  

In summary, both thermal and non-thermal processes are responsible for exciting the molecular hydrogen in the Bird and it is especially the emission from the outflowing gas contributing to the non-thermal excitation.  The Heart and Tail are dominated by thermal excitation. 

\begin{table}
  \centering
  \begin{tabular}{lcccccc}
  \hline
   \hline
Region   &                              Body & Body-nuc & Heart  & Heart-nuc & Head &  Tail   \\
 \hline
$T_{vib}$ [K]                 &  4900    &  3800   &   3900  &  3200  &  3800  &  3700   \\
$T_{rot}$  [K]                &  1700    &  2100   &   1300  &  2000  &  980  &  1200   \\
o / p                               &  3.5    &  3.2   &   3.0   &   2.4  &  2.7  &  2.6    \\
   \hline
Notes:           & flrscs.   &   shocks  &   shocks  &  shocks  &  UV  &  UV  \\
\hline
\end{tabular}
  \caption{\small Temperatures, and the ortho/para ratio, of the warm molecular gas derived from the SINFONI line-ratios (see text for details).   We also note the likely mechanism dominating the excitation of the H$_2$ gas in the relevant aperture, either thermal processes of shocks or UV photon heating, or non-thermal fluorescense.
   }
\label{temperatures}
\end{table}

\subsubsection{Diagnostic diagram}

\begin{figure}
\includegraphics[width=8.5cm,clip=true]{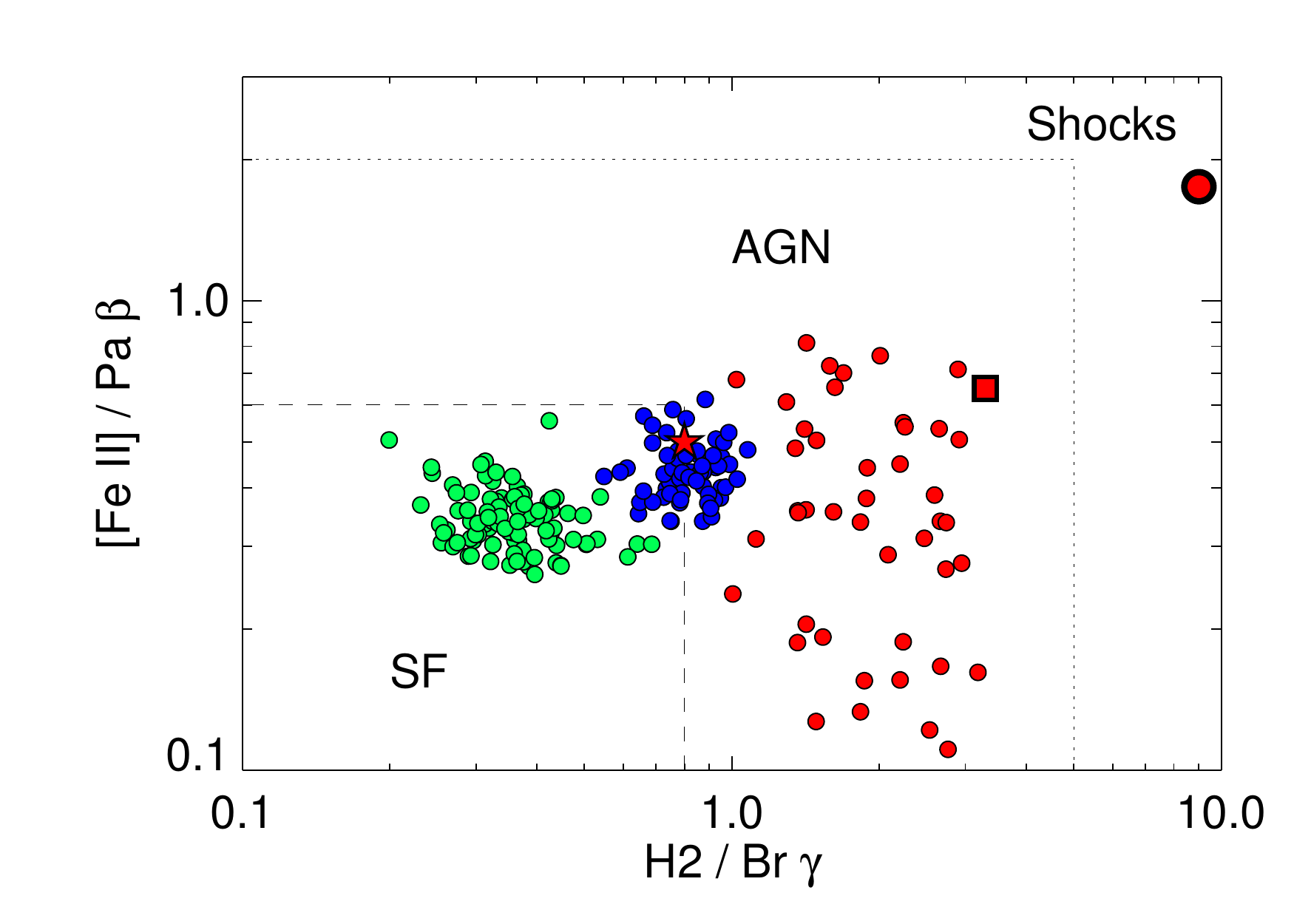}
\caption{\label{excitation} The [Fe II] / Pa$\beta$  ratio is shown as a function of H$_2$ 1-0 S(1) / Br$\gamma$, a diagnostic used to differentiate between sources of ionisation.  The coloured circles are those spaxels in the SINFONI data-cubes selected over the apertures of the Head (green), Heart (blue), and Body (red).  The different components separate remarkably well.  The values have been extinction corrected, though the effect is small due to the proximity of the lines. See text for discussion. The large red symbols show the Body H$_2$ / Br$\gamma$ ratios for the {\em velocity de-composed} nuclear aperture: blue-shifted gas (circle), systemic velocity gas (square), and redshifted gas (star) falling on top of the blue Heart points.}
\end{figure}

Strong emission line diagnostics can further help in disentangling between sources of excitation.
Figure~\ref{excitation} combines the [Fe II] / Pa$\beta$ ratio results vs. the H2 (1-0) S1 /  Br$\gamma$ ration {\em per spaxel} in the K250 data-cube.  Only the spaxels falling into the Head, and the nuclear regions of the Heart and Body are shown.  The two line ratios are corrected for extinction per spaxel, adopting values calculated from the strongest lines, Br$\gamma$ / Pa$\alpha$.  
The figure also shows another similar relation but using a Fe II / Br $\gamma$.  For this case, since the extinction correction become quite significant between the J and K-band datacubes, the per spaxel corrections becomes too noisy and we adopt the extinction correction using the 
A$_V$ (Br$\gamma$ / Pa$\alpha$) values in the Table~\ref{lineratios} for the three regions separately.
The three Bird components separate remarkably well in this diagnostic, which is used as an ionisation source indicator in a similar vein than BPT diagrams in the optical \citep{Riffel2013,Larkin1998}.  The diagram progresses from young star formation ionisation at lower left, through AGN related ionisation in the middle to shock ionisation at upper right, with recent SN activity showing up at high [Fe II] / Pa$\beta$ values.  We also constructed a similar figure using the [Fe II] / Br$\gamma$ ratio, as it is recently discussed in \citet{Colina2015}.  It shows very similar information as Fig.~\ref{excitation}, but is noisier due to the much larger extinction correction needed for in-between J and K band lines. 

The Head appears to be a prototypical young star-forming region where the excitation of gas is due to photo-ionisation.  The Heart falls at the boundary of the SF and AGN regions, though it is important to note that the spaxels specifically at the {\em nuclear} position are on the SF side of the line, while those surrounding the nucleus are more into the AGN area suggesting possible contamination from shocks, increasing the ratio slightly.  

In contrast, the red points corresponding to the Body aperture in Fig.~\ref{excitation} are squarely in the AGN region. However, the line ratios are still averaged over all the velocity components, and unfortunately it is not possible to reproduce the points per velocity component, the S/N in the spaxels does not allow it due to the faint Br$\gamma$ and Fe II, except for the central spaxel.  We determined the relevant line ratios separately for the decomposed velocities in the usual small Body aperture, however, and the striking result emerging is the very different ratios associated with the different velocities.  Figure~\ref{excitation} also shows these values: H$_2$ /  Br$\gamma$ is in the range $7-11$ for the blue-shifted gas (large red circle), it is $\approx3$ for the systemic velocity component (large red square), and $<1$ for the red-shifted component (red star).  The values are approximately the same for just the single central velocity decomposed spaxel of the Body.
The blue-shifted value falls in the "LINER" region of the diagnostic, suggesting primarily shock ionisation.  The redshifted component actually matches the velocity of the Head, and correspondingly shows a value close to photo-ionisation suggesting overlapping gas. 
The Body nuclear line ratios, however, remain in the AGN region.  This is interesting, since there are no other clear indications of AGN activity in the system.

\subsection{Is there an AGN hidden in the Bird?}

There were no indications of an AGN from the optical spectra, nor MIR diagnostics in the Bird system (V08).  However, it is quite plausible that optical diagnostics are masked by dust especially in the Body component, while MIR to FIR diagnostics using e.g.\ Spitzer and WISE data have too low spatial resolution to get anything but integrated values thus likely diluting any signal in case only one component has an AGN.  The suspicion naturally is directed to the Body, the most massive and evolved component of the galaxy system.
We analysed in more detail the highest S/N K-band spectrum extracted over the smallest Body aperture, in an attempt to isolate the nucleus as much as possible.  Analysis of this spectrum is complicated by the numerous H2 emission lines.  However, we see no sign of the coronal [SiVI] 1.9634 line in the spectrum even after carefully fitting and subtracting the neighbouring H$_2$ lines.  

The Bird is not detected in hard X-rays with BAT \citep{Koss2013}.  The Chandra GOALS survey on the other hand comfortably detects X-rays over the Bird system \citep{Iwasawa2011}, but the emission is soft, and extended, with the strongest concentration over the Head component, strongly suggesting SF rather than AGN origin.  They do note some ambiguities with the galaxy regarding its morphology and a possibly elevated X-ray [SiXIII] emission line. \citet{Yuan2010} reports the optical spectrum of the Bird as "composite" adding to the confusion.  However, these works are hampered by them considering Bird as a single object while it is clear from the present work that signals from different regions, if combined, make interpretations extremely difficult. 

We thus conclude that in the [Fe II] / Pa$\beta$ vs.\ H$_2$ (1-0)S1 /  Br$\gamma$ diagnostic, specifically the systemic velocity component of the small Body aperture (and also the central individual spaxel there) unambiguously lies in the AGN region defined in the literature.  While it still is difficult to absolutely rule out that the signal in this 300 pc spatial region would come from a mixture of photoionisation due to SF and shocks averaging out for typical AGN ratios, the simplest explanation (given that the shock signal is found preferentially in the outflow components and surrounding areas) is that indeed there is budding AGN activity at the location of the Body nucleus. This would not be surprising given the mass of the system.  As seen in the next Section, this is also the oldest SF area of the system, making it an interesting study of the onset of AGN signal after SF. 

Adopting the CO-based $\sigma$ value for the Body from Table~\ref{mainkinematics}, we estimate the mass of a central supermassive black hole (SMBH) as $M_{BH} \sim 10^{7.6} M_{\odot}$ \citep[e.g.][]{Gultekin2009}.  If an Eddington rate was assumed for accretion, the expected hard X-Ray luminosity would be above the detection level of BAT implying that (any) SMBHs in the Bird are currently accreting at sub-Eddington levels \citep[see e.g.][]{Merloni2003}.

\subsection{Age of recent star-formation in the Bird}
\label{sbage}

The H$_2$ /  Br $\gamma$ ratio (Fig.~\ref{excitation} and Table~\ref{lineratios}) is expected to increase with time, especially some 20 Myr after a starburst event due to rapidly decreasing Br$\gamma$ emission with disappearing OB-stars and growing H$_2$ emission with the related increase in SN activity and resulting winds shocking the gas.  Thus, the age of SF increases from component to component, the youngest ages appearing in the Head, then to the Tail and the Heart, and finally Body (see Table~\ref{eqwid}).  This is also consistent with the trend of the youngest stellar population ages from fits to the optical SALT spectra appearing in the Head (Fig.~\ref{sfh}).   
The Br$\gamma$ line EW values compared to SB99 based models \citep{Levesque2013} indicate ages of $\approx4-5$ Myr at the peak of emission in the Head, in its Eastern edge (see the Pa$\alpha$ and EW(Pa$\alpha$) maps in Figs.~\ref{featuremaps} and~\ref{ratiomaps}), and 6-7 Myr in the Head overall.  These time ranges are quite robust since the peak of the Head spectrum shows negligible CO band absorption (which is expected to appear rapidly at around 6-7 Myr), contrary to the Head as a whole, and the other regions, where the features are clear.  We checked the Br$\gamma$ EW's also against models presented in\citet{Puxley1997}, together with a CO-index, derived from the EW (CO 2.3$\mu$) as discussed therein and in \citet{Ryder2001}.  The Head values are consistent with the SB99 ages, while the Tail and Heart Br$\gamma$ and CO-index values suggests an age of $\approx$8 Myr for both of these.   {\em Within} the Head, the H$_2$/Br$\gamma$ ratio is larger in the outskirts. The Head [FeII]/Pa$\beta$ ratio is also larger in the outer parts compared to the inner regions (see Fig.~\ref{ratiomaps}, bottom left), consistent with the current SF being more compact than the somewhat older SF on the outer rim.  However, note that the recombination line EW maximum was rather at the eastern edge of the Head, as mentioned above, where the line-ratio maps already have low S/N.  
[He I] is thought to trace the very hottest and youngest stars, and indeed this emission is clearly strongest in the central regions of the Head (Fig.~\ref{featuremaps}), and interestingly also in the Tail. 

The Body is more complicated. The virtual non-existence of recombination emission at the nuclear location puts a lower limit for an age of an instantaneous starburst at around 8-9 Myr, when O-stars have all exploded.  The supernova rate, however, is expected to remain fairly constant until all $>8M_{\odot}$ stars have gone, which happens around 40 Myr based e.g.\ on SB99 models \citet{SB99}.  Since the [Fe II] emission is, at least partially, expected to be due to the SNe, and this emission is extremely weak here (Fig.~\ref{featuremaps}), we assign 40 Myr as a lower limit of the Body SF age.  We note that the \citet{Puxley1997} models would fit the Body values using a slow 20 Myr e-folding time starburst, and a time since the onset of such SF of more than 70 Myr.   Given that this is the only part in the Bird with any evidence of an AGN, such an apparent delay between a starburst and AGN activity fits well e.g.\ the findings of \citet[][]{Davies2007} from studies of Seyferts, and would make an interesting test of models of early AGN and starburst interplay. 

We also note that the age sequence of the components is the same as the progression of [Fe II] / Pa$\beta$ and H$_2$ (1-0)S1 /  B$\gamma$ line-ratios used in the diagnostic diagram in Fig.~\ref{excitation}, making the diagram potentially also an evolutionary stage diagnostic, perhaps excluding the pure shock region. 

The default Head aperture makes up approximately 45\% of the Bird Pa$\alpha$ emission in the SINFONI field of view including all the diffuse outlying emission, and more than 60\% of the summed up ionised flux from the apertures analysed previously.  Since it is the Head Pa$\alpha$ that was found to be purely SF related, and the emission in the Heart and Body regions to be mixed with shock heating, we assign a {\em lower} limit of 60 \% of the total  SFR originating from the Head.  Broadly consistent with this, an approximate spatial decomposition of the Spitzer 24$\mu$m map (V08) shows some 60-70\% of SFR peaking in the Head area.  

\begin{figure}
\includegraphics[width=8cm,clip=true]{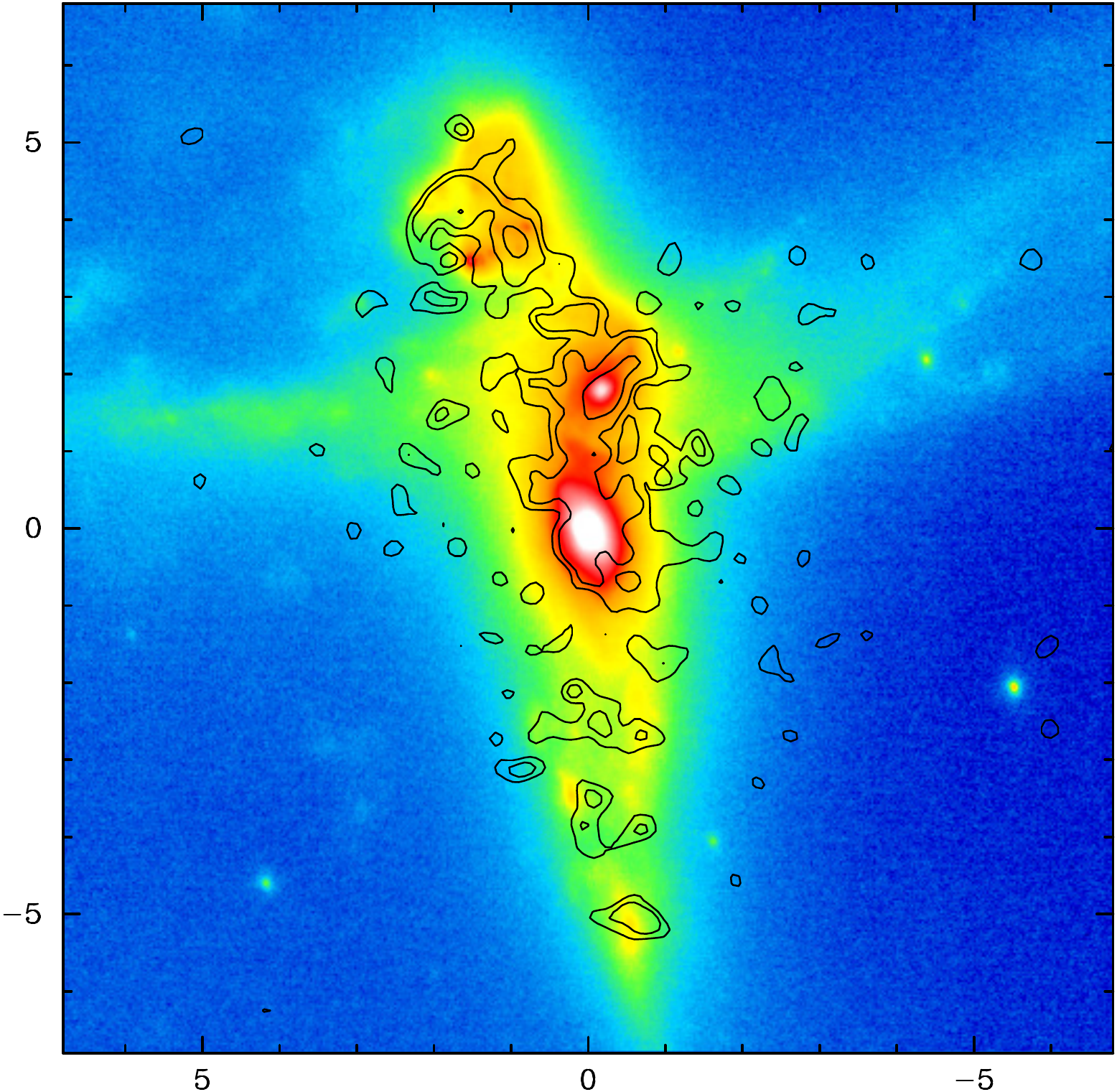}
\caption{\label{visirpic}  The 11.3 $\mu$m PAH emission from VISIR observations is shown as contours overlaid on the $K$-band adaptive optics image from V08. 
The strongest PAH comes from the location of the Heart component, while it appears depleted compared to SF in the Head areas.  
 }
\end{figure}

The VISIR data of the Bird are shown in Fig.~\ref{visirpic}) as countours, overlaid on the K-band continuum data.  While there is PAH flux from the Head, the strongest PAH emission comes from the Heart.  PAH appears to avoid the very strongest SF regions, as we have found in another LIRG, NGC~1614 \citep{Vaisanen2012}.   PAH emission may be depleted in the Head area due to the destruction of PAH carriers or e.g.\ due to dilution by hotter dust continuum, which would be consistent with very red Spitzer/IRAC $3.6 - 4.5 \ \mu$m colour of the Head (see V08), typical to dwarf galaxies  \citep{Smith2009}, though the Head is by no means a dwarf. 
Based on the ages determined above for the various Bird components, we can say that the conditions for PAH destruction or dilution happen at starburst ages of $<7$ Myr.

Finally, we note that at narrow velocity slices of Pa$\alpha$ emission, around 14810 to 14900 km/s, we identify several point-like knots in the Heart and especially Head areas.  These could be very young Super Star Clusters, or Young Massive Clusters, or at least complexes thereof, as is often seen in nearby gas-rich galaxy mergers or other intense massive star-formation regions \citep[][and references therein]{Randriamanakoto2013,Vaisanen2014}.
K-band adaptive optics imaging showed a number of point-like sources in this galaxy, concentrating on the side facing the larger galaxy. 
This distribution of star formation here is in agreement with triggering of formation of massive clusters by compression of gas \citep[e.g.][]{Renaud2015}. The S/N of the present data is not sufficient, however, to study the spectra of these features independently.

\subsection{Histories of the Bird components}
\label{histories}

In Section~\ref{stellarpops} we found that the Body+Heart region appears to have a strongly bimodal star formation history, with significant populations of stars formed some $\sim1$ Gyr ago, as well as recently  $\sim10$ Myr ago (see Fig.~\ref{sfh}).  The Tail has only the younger peak, and both the Tail and the Head appear to have more evenly distributed SFH overall.  In general, the stellar population results in the Bird are not unlike those in LIRG stellar population studies using wide wavelength range SED-fitting techniques \citep[e.g.][]{PereiraSantaella2015}, often showing strong recent star-formation on top of an otherwise older evolved stellar-population.   
The age of the older 1 Gyr stellar population is consistent with a scenario of elevated star-formation due to an earlier encounter of the Body and Heart galaxies at their (likely) first approach.  Typical timescales between first passage and coalescence ion mergers producing strong tidal tails are of this order \citep[e.g.][]{Khockfar2006,Peter2009}. It is furthermore tempting to speculate that the more continuous star formation in the Head is a tell-tale sign that it was not involved in the Bird interaction then, but has only joined more recently with the youngest starburst.  Whatever its history, the cause of the strongest young $<6$ Myr starburst in the Head clearly has to be due to the on-going interaction.  Whether the SF in the other components are due to the same interaction, or rather due to the Heart and Body (presumably) coming together in their second approach, is more difficult to say.  The older than $>40$ Myr starburst in the Body nucleus may suggest that the second approach has happened 50-100 Myr ago, whereas the Heart appears to be more affected by the on-going interaction with the Head, as seen e.g. in the shocked regions in between those components.

\subsubsection{Two or three galaxies?}
\label{kinedisc}

Apart from dynamical modelling which is out of the scope of this paper, are there any other observational clues to the origin of the Head?
The Oxygen abundances are uniform throughout the system (O/H $\approx 8.8$; Table~\ref{slfits}), and are close to, or just below, to what is expected for our $\sim 5-10 \times 10^{10} \, M_{\odot}$ stellar mass galaxies \citep[e.g.][]{Kewley2008}.  This can easily be understood by inflows of and mixing of ionised gas in the ongoing interaction \citep[e.g.][]{Rich2012,Perez2011}, where the likely interaction time scale would allow enough time for enriched gas to spread widely.  

The lower stellar population [Fe/H] abundance of the Head is consistent with the scenario of its separate origin, in that neither the mixing nor the recent enrichment would yet have affected the older stellar population, and the 0.2 dex lower [Fe/H] matches what is expected from the mass difference.  However, a similarly lower [Fe/H] could also be expected if the Head was a remote part of an progenitor disk involving the Body -- stellar metallicity gradient determinations \citep[e.g.][]{Pilkington2012,Sanchez2014} are quite noisy, however, and moreover, in that case the Tail would be expected to have a similar lower value, which it does not have.

The general kinematics were presented in Section~\ref{kinres} and Fig.~\ref{pa_vel}, consisting of the horizontal Heart-centred rotation, and the more complex vertical structure including the disjointed Head velocities.
The new information from the NIR spectra in this work was that the Body systemic velocity is $\sim$14750 km/s rather than the 14600 km/s measured from optical data in V08, and also that the Body nucleus actually appears to be encircled by a rapidly rotating gas disk.  The Body systemic velocity is in between the Head and the Tail radial velocities, raising the possibility that the north-south structure of the Bird system could be interpreted as a single large progenitor galaxy, with the Body as the centre (and the horizontal Heart progenitor in the foreground hiding the details in between). We show a pseudo-longslit 2D spectrum from the K250 data-cube, more or less corresponding to the SALT spectrum in Fig.~\ref{saltspec}.  Though the Head and Tail are quite asymmetric about the Body (which also shows the rotation disk discussed above), the Head does appear to be connected to the Body with a faint Pa$\alpha$ velocity structure, making the two-galaxy scenario plausible.  However, the connecting material could easily be tidal material from either one.  Furthermore,  we know from the velocity field in Fig.~\ref{pa_vel} that the Head rotates in a different axis than the Tail/Body, the maximum rotation of the Head is achieved along PA$\approx -20$ rather than along the cut pictured at PA=15.   Also, the Head produces significant continuum light, while the Tail's is significantly weaker, and the stellar metallicity was found to be lower in the Head, compared to the Heart, Body, and Tail.  

In summary, we argue that a three-galaxy interpretation is the simplest one, matching the bulk of the observations. The field is very complex, however, and an ultimate verdict is difficult to make before proper simultaneous rotating disk modelling of at least two and three individual distorted disks.  We finally note that in case there was only one vertical progenitor galaxy, it would be an extremely massive one, with uncorrected rotational velocities more than 300 km/s and a diameter in excess of 30 kpc, with an admittedly very intriguing ULIRG-level starburst happening in {\em one half} of the progenitor.

\begin{figure}
\includegraphics[width=8.5cm,clip=true]{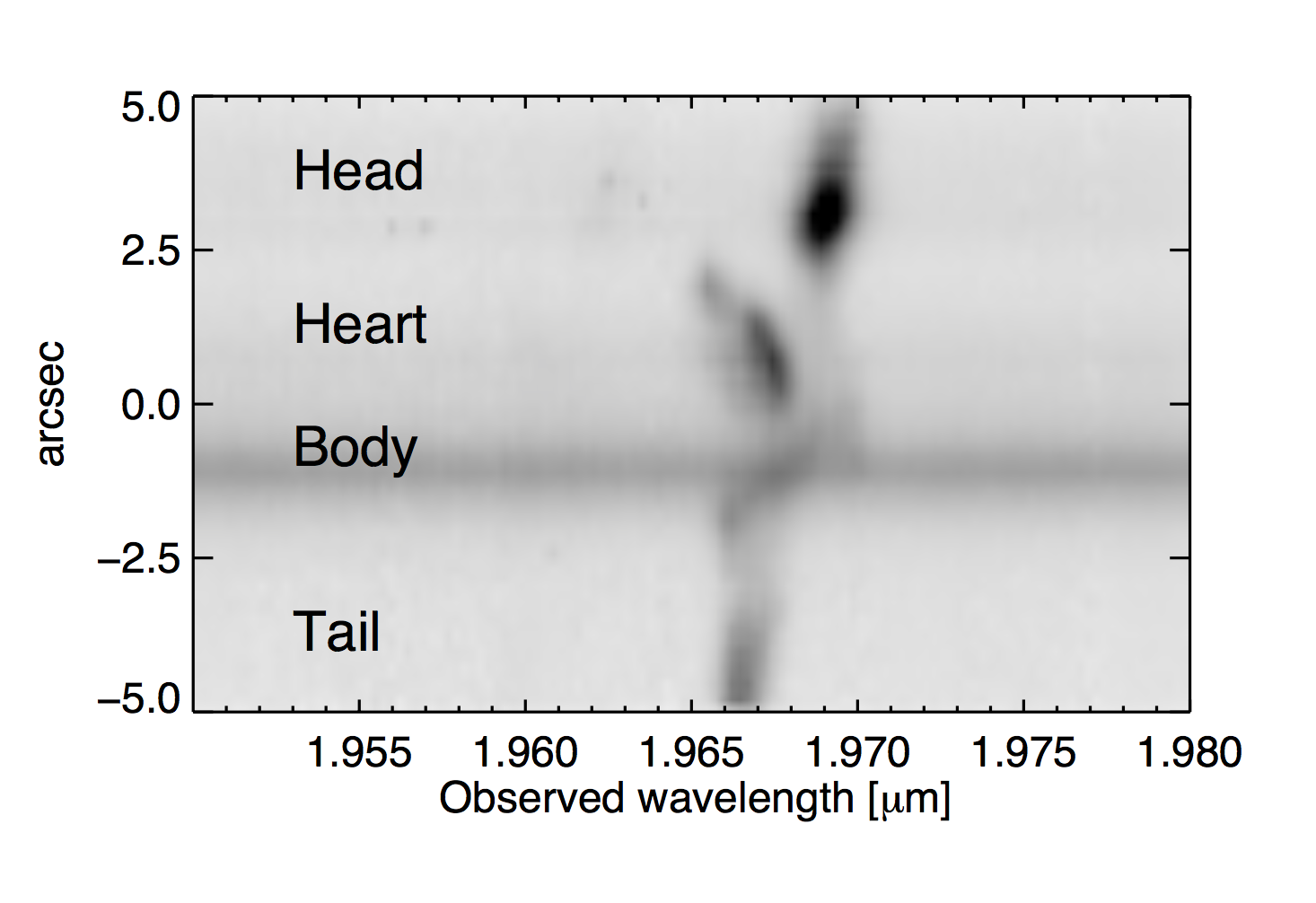}
\caption{\label{pseudoslit}  A pseudo-longslit at PA=15 constructed from the SINFONI K250 data-cube, showing the Pa$\alpha$ location.  The main features are marked. The strongest continuum is evident at the Body location. The spectrum can be compared to the 2D SALT long-slit spectrum in Fig.~\ref{saltspec}.  
}
\end{figure}

\subsubsection{Off-nuclear starburst and multiple nuclei}

Regardless of the kinematic scenario adopted, the Bird presents a spectacular off-nuclear starburst, and thus an example of a (U)LIRGs which does not easily conform to the standard picture of a powerful central starburst.  Other such off-nuclear starburst examples include II Zw 096 \citep{Inami2010}, and the closest and very well-studied gas-rich merger system, the Antennae, where the strongest starburst is in an overlap region in-between the merging nuclei \citep[e.g.][]{Mirabel1998,Karl2010}, and the Tadpole galaxy \citep{Jarrett2006} where the strongest SF is in the main disk in a possible ring structure.  Herrero-Illana et al. (2017, subm.) also found an off-nucleus starburst visible only in the MIR in IRAS~16516-0948. 
In case of (U)LIRGs a complicating factor is that some nuclei of progenitor systems can be so obscured that they are missed in analyses \citep[e.g.][]{Haan2011, Inami2010}, as indeed is the case of our galaxy. 

That the Head is forming stars is not surprising in itself:  estimating the Toomre parameter \citep{toomre,Westmoquette2012} from the kinematics in Table~\ref{mainkinematics}, we get $Q \approx 0.5$, i.e.\ the gas disk is unstable, and prone to collapse. The Heart also has $Q<1$, whereas the parameter in the Body is slightly above or below unity, depending on which $\sigma$ value is used. 

Note that there may be a connection of off-nuclear starbursts to multiple merging \citep[e.g.][]{Borne2000}, and/or the possibility these systems are evidence that ULIRGs typically happen in (compact) groups of galaxies \citep[e.g.][]{Amram2007}.  Indeed, it has been shown that (U)LIRGs tend to favour the group-size halo masses \citep{Tekola2014}, which would naturally explain why (U)LIRGs happen in mergers of more than two progenitors.  Tight groups of galaxies, such as Hickson compact groups, may well be candidates for their future evolution playing out similarly to the Bird system \citep[see e.g.][]{Borne2000}.

\section{Bird at high-redshift}

\begin{figure}
\includegraphics[width=4cm,clip=true]{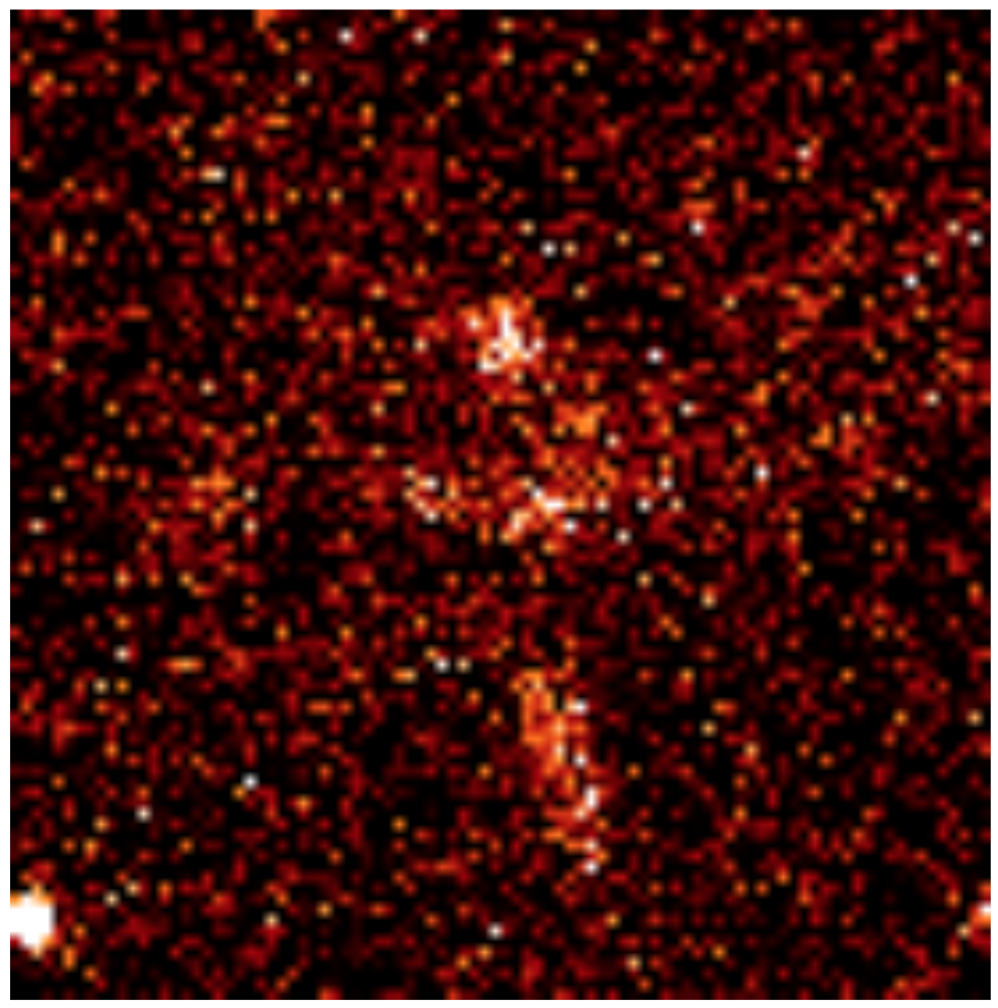}
\includegraphics[width=4cm,clip=true]{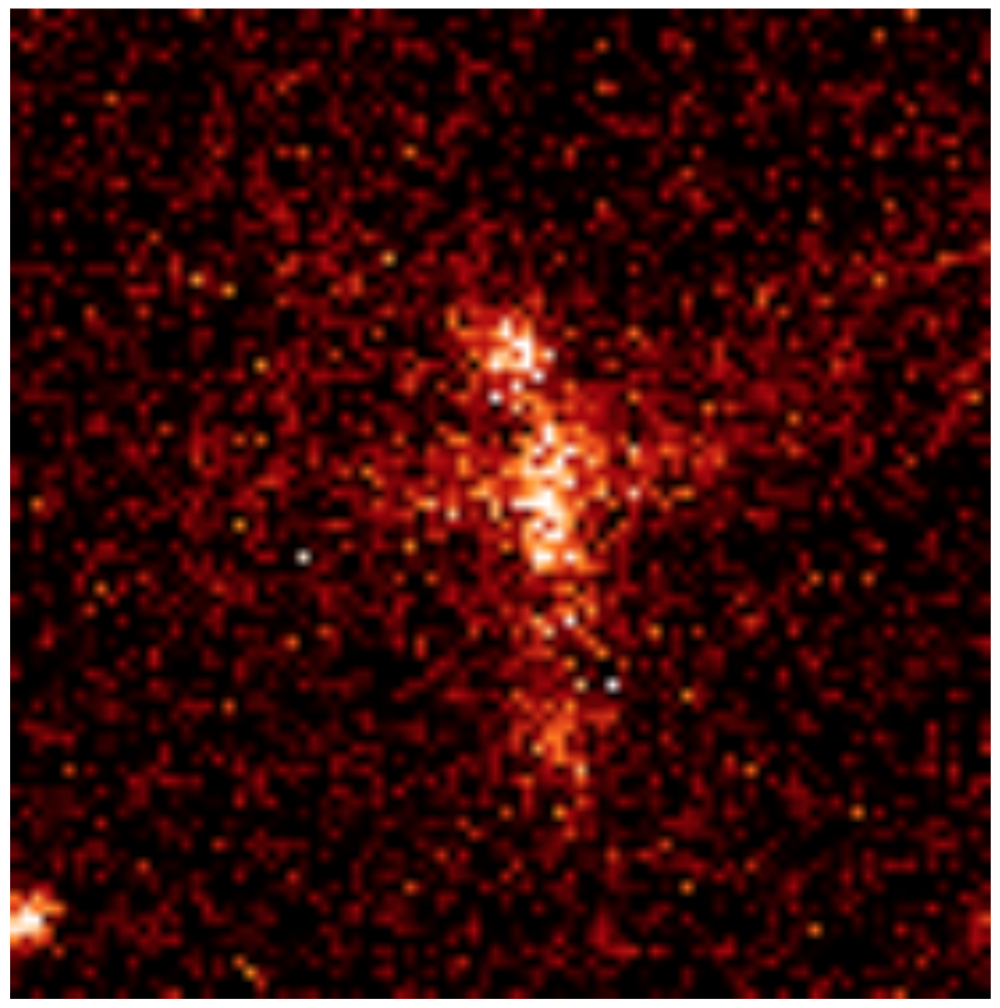}
\caption{\label{redshift}  Bird as seen from redshifts $z \sim 1-2$ with 10-m class AO-instruments or the HST.  The left panel is based on rest-frame B-band Bird image, and corresponds to an J-band observation of a $z=2$ target, or an I-band observation of a $z=1$ galaxy.  The right panel is based on the I-band image, and corresponds to e.g.\ a $z=2$ galaxy observed in K-band.  The simulated size of the image is 2.5", or about 20 kpc in the $z=1-2$ range.
}
\end{figure}

What would the Bird look like at high redshift?   As an example, Fig.~\ref{redshift} shows the HST B and I-band images (V08) convolved, re-binned, and dimmed to scales expected at $z = 1-2$ where 1\arcsec corresponds to approximately 9 kpc.   The background noise is scaled assuming a similar depth observation as the original HST images, in between 800 and 1200 sec. The images still assume HST or 10-m class telescope adaptive optics resolution and relevant pixel-scales, $~\sim 0.1$" and 0.03"/px, respectively.   At $z \approx 2$ an observed J-band image would be looking at rest-frame B (the left panel), and K-band would see the I-image (the right panel).   At $z \approx 1$ an I-band observation would see the rest-frame B-band, while an observed J-band would include H$\alpha$, and perhaps look even more extreme with the Head dominating (cf.\ the Pa$\alpha$ panel in Fig.~\ref{featuremaps}).

The bottom line is that high-$z$ red/infrared observations would detect a very {\em clumpy} galaxy \citep[e.g.][]{Shibuya2016} , especially any observations probing the rest-frame I-band and likely also V-bands.   
Observations detecting wavelengths close to the rest-frame B-band would, on the other hand, likely interpret the system as a tight group of compact $\sim 1$ kpc size galaxies (left panel of Fig.~\ref{redshift}). 
Optical observations would be looking at the UV view of the Bird, and would see only see the Head, and perhaps some outlying fuzz from the wings and some parts of the Tail.  The Head would look like a strongly star-forming compact galaxy.  The Body would be totally invisible in all of these, as it already is in the zero-redshift B-band image.  

Would the velocity fields be recovered?   If redshifted H$\beta$ or H$\alpha$-were available with NIR instrumnts, the Head, Heart and Tail components would be recovered, but they would entirely miss the true Body component and its significant rapidly rotating gas disk, as already happened with our optical SALT observations (see Section~\ref{kinres}).  Head would show a bit of rotation, and the tidal tails might show huge velocity offsets relative to it.   These would likely be interpreted as high-velocity winds, though in reality they are components of the interacting/merging system.

Hence, we argue, that careful studies of samples of local (U)LIRGs are needed to interpret observations and results from higher redshift galaxy surveys.  In particular, integrated properties can be quite mis-leading, in case closely interacting or merging systems are in significantly different evolutionary stages, as found in the case of the Bird.

\section{Summary}

We have presented a detailed study of a local interacting (U)LIRG, IRAS~19115-2124, aka the Bird, using new VLT/SINFONI NIR integral field spectroscopy and optical SALT/RSS long-slit spectroscopy.   This has been done to understand the past history of a merging gas-rich galaxy system and study in detail the physical mechanisms that transform galaxy populations on cosmic timescales, namely the triggering and evolution of star formation, its quenching, as well as the onset of active galaxy nuclei, and how galaxy interactions affect these.  We also wanted to see how this target would be interpreted at high redshift, to help understand $z=1-2$ ULIRGs.  The main results can be summarised as follows:

We confirm the suggestion based on our earlier adaptive optics imaging \citep{Vaisanen2008}, that the dominant source of the large IR luminosity, $log(L_{IR} / L_{\odot}) \approx 11.9$, is a small irregular looking component of the interaction 3 kpc away from two primary galaxy nuclei.  Based on the recombination line emission maps, we find that this component, the Head of the Bird, produces more than 60\% of the total IR, and SFR, of the whole galaxy system, also containing the Heart and Body components.  

We derive rotational velocities and measure velocity dispersions from both emission lines and the CO band-head absorption for the distinct Bird components.  The Heart and Head appear rotationally supported, while the Body has the highest velocity dispersions. The dynamical masses are estimated to be approximately $1 \times 10^{11} M_{\odot}$,  $7 \times 10^{10} M_{\odot}$, and $2 \times 10^{10} M_{\odot}$ for the Body, Heart, and Head, respectively.  We also find in the new data a rapidly rotating inner disk in the Body.  
 
Excitation derived from NIR line-ratios is shown to be a mix of thermal and non-thermal mechanisms, with the Body nucleus especially contributing to the latter.  While shock heating is present everywhere, the Head and Tail appear to be dominated by UV photon heating.  The three main components separate very well in a [Fe II] / Pa$\beta$ vs.\ H$_2$ (1-0)S1 /  B$\gamma$ diagnostic diagram.  These results show the Head to be a pure starburst, while the Heart emission is characteristic of star formation and shock heating signal.  The Body is a complex case, but after investigating the line-ratios of decoupled velocity components, we interpret its shock heating to come from gas outflows, while the systemic and nuclear signal is consistent with an existence of an AGN. The Body nucleus appears devoid of current star formation.

A major goal of this study was to derive the sequence of star formation in the various Bird components.  The NIR line-ratios, together with EWs, and optical stellar population fitting to the SALT spectra, reveal the ages of the most recent starbursts as well as the older underlying populations.  The Head has very young 4--7 Myr SF, while the Heart, and the Tail (essentially the tidal tail of the Body), are slightly older at 8 Myr age.   The Body, in contrast, has apparently stopped forming stars at least 40 Myr ago.  It is thus an example of a very {\em recently quenched galaxy nucleus}, with budding AGN activity.  The younger regions do not show any signs of AGN activity, suggesting a delay of at least $\sim40$ Myr between a starburst and significant feeding of a central supermassive black hole in this case.  

Stellar population history fits suggest a major star-forming episode approximately 1 Gyr ago, plausibly during a first encounter of the main two nuclei.  A slightly different SFH fit to the Bird spectrum, and its lower stellar metallicity, [Fe/H] $\approx 0.011$ vs.\ close to Solar elsewhere, is evidence, apart from the kinematics, that the Head might have had a different history.  We speculate it has joined the interaction later, though this would need to be modelled. The gas-phase metallicities on the other hand are uniform around the system suggesting rapid gas flows and mixing within a few tens of Myr during the latest interaction and star formation episode.

The Bird interaction shows gas flows in multiple spatial locations and in different gas phases.  These are studied in detail in a follow-up work (V\"ais\"anen et al., in prep), correlating their characteristics to the different starburst timescales over the Bird components uncovered in this work.  We also have recently obtained ALMA observations of this target to study the bulk of the molecular gas reservoirs aiding an attempt to describe the future of this galaxy system (Romero-Canizales et al., in prep.).

Finally, we show what the Bird would look like at high redshift.  All integrated properties of the system would be hopelessly misguided, due to them originating from nuclei and regions with such different evolutionary stages, ranging from a current young starburst to a quenched nucleus. Depending on filter used, in a good case it would look like a very clumpy galaxy, and in a worse case it would easily be mistaken for a group of compact galaxies.  For example the massive nucleus of the Body with its significant rotating inner disk would likely be missed altogether.

\section*{Acknowledgements}

PV acknowledges support from the National Research Foundation of South Africa and FINCA.  P.H.J. acknowledges the support of the Academy of Finland grant 1274931.  Results presented in this paper are based on observations made with ESO Telescopes at the Paranal Observatory under programs 383.B-0637(B) and 385.B-0712(A), and the Southern African Large Telescope (SALT) under programs 2011-2-RSA\_OTH-002 and 2011-3-RSA\_OTH-023. 

\bibliographystyle{mn2e}

\bibliography{bird_merger}

\begin{thebibliography}{69}
\expandafter\ifx\csname natexlab\endcsname\relax\def\natexlab#1{#1}\fi

\bibitem[{{Alatalo} {et~al}\mbox{.}(2016){Alatalo}, {Cales}, {Rich},
  {Appleton}, {Kewley}, {Lacy}, {Lanz}, {Medling}, \& {Nyland}}]{Alatalo2016}
{Alatalo} K. {et~al.}, 2016, \apjs, 224, 38

\bibitem[{{Amram} {et~al}\mbox{.}(2007){Amram}, {Mendes de Oliveira}, {Plana},
  {Balkowski}, \& {Hernandez}}]{Amram2007}
{Amram} P., {Mendes de Oliveira} C., {Plana} H., {Balkowski} C., {Hernandez}
  O., 2007, \aap, 471, 753

\bibitem[{{Barnes} \& {Hernquist}(1996)}]{Barnes1996}
{Barnes} J.~E., {Hernquist} L., 1996, \apj, 471, 115

\bibitem[{{Borne} {et~al}\mbox{.}(2000){Borne}, {Bushouse}, {Lucas}, \&
  {Colina}}]{Borne2000}
{Borne} K.~D., {Bushouse} H., {Lucas} R.~A., {Colina} L., 2000, \apjl, 529, L77

\bibitem[{{Bruzual} \& {Charlot}(2003)}]{BC03}
{Bruzual} G., {Charlot} S., 2003, \mnras, 344, 1000

\bibitem[{{Burgh} {et~al}\mbox{.}(2003){Burgh}, {Nordsieck}, {Kobulnicky},
  {Williams}, {O'Donoghue}, {Smith}, \& {Percival}}]{Burgh2003}
{Burgh} E.~B., {Nordsieck} K.~H., {Kobulnicky} H.~A., {Williams} T.~B.,
  {O'Donoghue} D., {Smith} M.~P., {Percival} J.~W., 2003, in Proc.\ SPIE, Vol.
  4841, Instrument Design and Performance for Optical/Infrared Ground-based
  Telescopes, {Iye} M., {Moorwood} A.~F.~M., eds., pp. 1463--1471

\bibitem[{{Cappellari}(2016)}]{ppxf}
{Cappellari} M., 2016, ArXiv e-prints

\bibitem[{{Cid Fernandes} {et~al}\mbox{.}(2005){Cid Fernandes}, {Mateus},
  {Sodr{\'e}}, {Stasi{\'n}ska}, \& {Gomes}}]{CidFernandes2005}
{Cid Fernandes} R., {Mateus} A., {Sodr{\'e}} L., {Stasi{\'n}ska} G., {Gomes}
  J.~M., 2005, \mnras, 358, 363

\bibitem[{{Colina} {et~al}\mbox{.}(2015){Colina}, {Piqueras L{\'o}pez},
  {Arribas}, {Riffel}, {Riffel}, {Rodriguez-Ardila}, {Pastoriza},
  {Storchi-Bergmann}, {Alonso-Herrero}, \& {Sales}}]{Colina2015}
{Colina} L. {et~al.}, 2015, \aap, 578, A48

\bibitem[{{Crawford} {et~al}\mbox{.}(2010){Crawford}, {Still}, {Schellart},
  {Balona}, {Buckley}, {Dugmore}, {Gulbis}, {Kniazev}, {Kotze}, {Loaring},
  {Nordsieck}, {Pickering}, {Potter}, {Romero Colmenero}, {Vaisanen},
  {Williams}, \& {Zietsman}}]{Crawford2010}
{Crawford} S.~M. {et~al.}, 2010, in Proc.\ SPIE, Vol. 7737, Observatory
  Operations: Strategies, Processes, and Systems III, p. 773725

\bibitem[{{Davies} {et~al}\mbox{.}(2007){Davies}, {M{\"u}ller S{\'a}nchez},
  {Genzel}, {Tacconi}, {Hicks}, {Friedrich}, \& {Sternberg}}]{Davies2007}
{Davies} R.~I., {M{\"u}ller S{\'a}nchez} F., {Genzel} R., {Tacconi} L.~J.,
  {Hicks} E.~K.~S., {Friedrich} S., {Sternberg} A., 2007, \apj, 671, 1388

\bibitem[{{Eisenhauer} {et~al}\mbox{.}(2003){Eisenhauer}, {Abuter}, {Bickert},
  {Biancat-Marchet}, {Bonnet}, {Brynnel}, {Conzelmann}, {Delabre}, {Donaldson},
  {Farinato}, {Fedrigo}, {Genzel}, {Hubin}, {Iserlohe}, {Kasper},
  {Kissler-Patig}, {Monnet}, {Roehrle}, {Schreiber}, {Stroebele}, {Tecza},
  {Thatte}, \& {Weisz}}]{Eisenhauer2003}
{Eisenhauer} F. {et~al.}, 2003, in Proc.\ SPIE, Vol. 4841, Instrument Design
  and Performance for Optical/Infrared Ground-based Telescopes, {Iye} M.,
  {Moorwood} A.~F.~M., eds., pp. 1548--1561

\bibitem[{{Emonts} {et~al}\mbox{.}(2014){Emonts}, {Piqueras-L{\'o}pez},
  {Colina}, {Arribas}, {Villar-Mart{\'{\i}}n}, {Pereira-Santaella},
  {Garcia-Burillo}, \& {Alonso-Herrero}}]{Emonts2014}
{Emonts} B.~H.~C., {Piqueras-L{\'o}pez} J., {Colina} L., {Arribas} S.,
  {Villar-Mart{\'{\i}}n} M., {Pereira-Santaella} M., {Garcia-Burillo} S.,
  {Alonso-Herrero} A., 2014, \aap, 572, A40

\bibitem[{{Freudling} {et~al}\mbox{.}(2013){Freudling}, {Romaniello},
  {Bramich}, {Ballester}, {Forchi}, {Garc{\'{\i}}a-Dabl{\'o}}, {Moehler}, \&
  {Neeser}}]{Freudling2013}
{Freudling} W., {Romaniello} M., {Bramich} D.~M., {Ballester} P., {Forchi} V.,
  {Garc{\'{\i}}a-Dabl{\'o}} C.~E., {Moehler} S., {Neeser} M.~J., 2013, \aap,
  559, A96

\bibitem[{{Gallazzi} {et~al}\mbox{.}(2005){Gallazzi}, {Charlot}, {Brinchmann},
  {White}, \& {Tremonti}}]{Gallazzi2005}
{Gallazzi} A., {Charlot} S., {Brinchmann} J., {White} S.~D.~M., {Tremonti}
  C.~A., 2005, \mnras, 362, 41

\bibitem[{{G{\"u}ltekin} {et~al}\mbox{.}(2009){G{\"u}ltekin}, {Richstone},
  {Gebhardt}, {Lauer}, {Tremaine}, {Aller}, {Bender}, {Dressler}, {Faber},
  {Filippenko}, {Green}, {Ho}, {Kormendy}, {Magorrian}, {Pinkney}, \&
  {Siopis}}]{Gultekin2009}
{G{\"u}ltekin} K. {et~al.}, 2009, \apj, 698, 198

\bibitem[{{Haan} {et~al}\mbox{.}(2011){Haan}, {Surace}, {Armus}, {Evans},
  {Howell}, {Mazzarella}, {Kim}, {Vavilkin}, {Inami}, {Sanders}, {Petric},
  {Bridge}, {Melbourne}, {Charmandaris}, {Diaz-Santos}, {Murphy}, {U},
  {Stierwalt}, \& {Marshall}}]{Haan2011}
{Haan} S. {et~al.}, 2011, \aj, 141, 100

\bibitem[{{Hopkins} {et~al}\mbox{.}(2006){Hopkins}, {Hernquist}, {Cox}, {Di
  Matteo}, {Robertson}, \& {Springel}}]{Hopkins2006}
{Hopkins} P.~F., {Hernquist} L., {Cox} T.~J., {Di Matteo} T., {Robertson} B.,
  {Springel} V., 2006, \apjs, 163, 1

\bibitem[{{Inami} {et~al}\mbox{.}(2010){Inami}, {Armus}, {Surace},
  {Mazzarella}, {Evans}, {Sanders}, {Howell}, {Petric}, {Vavilkin}, {Iwasawa},
  {Haan}, {Murphy}, {Stierwalt}, {Appleton}, {Barnes}, {Bothun}, {Bridge},
  {Chan}, {Charmandaris}, {Frayer}, {Kewley}, {Kim}, {Lord}, {Madore},
  {Marshall}, {Matsuhara}, {Melbourne}, {Rich}, {Schulz}, {Spoon}, {Sturm},
  {U}, {Veilleux}, \& {Xu}}]{Inami2010}
{Inami} H. {et~al.}, 2010, \aj, 140, 63

\bibitem[{{Iwasawa} {et~al}\mbox{.}(2011){Iwasawa}, {Sanders}, {Teng}, {U},
  {Armus}, {Evans}, {Howell}, {Komossa}, {Mazzarella}, {Petric}, {Surace},
  {Vavilkin}, {Veilleux}, \& {Trentham}}]{Iwasawa2011}
{Iwasawa} K. {et~al.}, 2011, \aap, 529, A106

\bibitem[{{Jarrett} {et~al}\mbox{.}(2006){Jarrett}, {Polletta}, {Fournon},
  {Stacey}, {Xu}, {Siana}, {Farrah}, {Berta}, {Hatziminaoglou}, {Rodighiero},
  {Surace}, {Domingue}, {Shupe}, {Fang}, {Lonsdale}, {Oliver},
  {Rowan-Robinson}, {Smith}, {Babbedge}, {Gonzalez-Solares}, {Masci},
  {Franceschini}, \& {Padgett}}]{Jarrett2006}
{Jarrett} T.~H. {et~al.}, 2006, \aj, 131, 261

\bibitem[{{Johansson} {et~al}\mbox{.}(2009){Johansson}, {Naab}, \&
  {Burkert}}]{Peter2009}
{Johansson} P.~H., {Naab} T., {Burkert} A., 2009, \apj, 690, 802

\bibitem[{{Karl} {et~al}\mbox{.}(2010){Karl}, {Naab}, {Johansson}, {Kotarba},
  {Boily}, {Renaud}, \& {Theis}}]{Karl2010}
{Karl} S.~J., {Naab} T., {Johansson} P.~H., {Kotarba} H., {Boily} C.~M.,
  {Renaud} F., {Theis} C., 2010, \apjl, 715, L88

\bibitem[{{Kauffmann} {et~al}\mbox{.}(2003){Kauffmann}, {Heckman}, {White},
  {Charlot}, {Tremonti}, {Brinchmann}, {Bruzual}, {Peng}, {Seibert},
  {Bernardi}, {Blanton}, {Brinkmann}, {Castander}, {Cs{\'a}bai}, {Fukugita},
  {Ivezic}, {Munn}, {Nichol}, {Padmanabhan}, {Thakar}, {Weinberg}, \&
  {York}}]{Kauffmann2003}
{Kauffmann} G. {et~al.}, 2003, \mnras, 341, 33

\bibitem[{{Kewley} \& {Ellison}(2008)}]{Kewley2008}
{Kewley} L.~J., {Ellison} S.~L., 2008, \apj, 681, 1183

\bibitem[{{Khochfar} \& {Burkert}(2006)}]{Khockfar2006}
{Khochfar} S., {Burkert} A., 2006, \aap, 445, 403

\bibitem[{{Koss} {et~al}\mbox{.}(2013){Koss}, {Mushotzky}, {Baumgartner},
  {Veilleux}, {Tueller}, {Markwardt}, \& {Casey}}]{Koss2013}
{Koss} M., {Mushotzky} R., {Baumgartner} W., {Veilleux} S., {Tueller} J.,
  {Markwardt} C., {Casey} C.~M., 2013, \apjl, 765, L26

\bibitem[{{Krajnovi{\'c}} {et~al}\mbox{.}(2006){Krajnovi{\'c}}, {Cappellari},
  {de Zeeuw}, \& {Copin}}]{kinemetry2006}
{Krajnovi{\'c}} D., {Cappellari} M., {de Zeeuw} P.~T., {Copin} Y., 2006,
  \mnras, 366, 787

\bibitem[{{Lagage} {et~al}\mbox{.}(2004){Lagage}, {Pel}, {Authier}, {Belorgey},
  {Claret}, {Doucet}, {Dubreuil}, {Durand}, {Elswijk}, {Girardot}, {K{\"a}ufl},
  {Kroes}, {Lortholary}, {Lussignol}, {Marchesi}, {Pantin}, {Peletier},
  {Pirard}, {Pragt}, {Rio}, {Schoenmaker}, {Siebenmorgen}, {Silber}, {Smette},
  {Sterzik}, \& {Veyssiere}}]{Lagage2004}
{Lagage} P.~O. {et~al.}, 2004, The Messenger, 117, 12

\bibitem[{{Larkin} {et~al}\mbox{.}(1998){Larkin}, {Armus}, {Knop}, {Soifer}, \&
  {Matthews}}]{Larkin1998}
{Larkin} J.~E., {Armus} L., {Knop} R.~A., {Soifer} B.~T., {Matthews} K., 1998,
  \apjs, 114, 59

\bibitem[{{Larson} {et~al}\mbox{.}(2016){Larson}, {Sanders}, {Barnes},
  {Ishida}, {Evans}, {U}, {Mazzarella}, {Kim}, {Privon}, {Mirabel}, \&
  {Flewelling}}]{Larson2016}
{Larson} K.~L. {et~al.}, 2016, \apj, 825, 128

\bibitem[{{Leitherer} {et~al}\mbox{.}(1999){Leitherer}, {Schaerer}, {Goldader},
  {Delgado}, {Robert}, {Kune}, {de Mello}, {Devost}, \& {Heckman}}]{SB99}
{Leitherer} C. {et~al.}, 1999, \apjs, 123, 3

\bibitem[{{Levesque} \& {Leitherer}(2013)}]{Levesque2013}
{Levesque} E.~M., {Leitherer} C., 2013, \apj, 779, 170

\bibitem[{{Maraston}(2005)}]{Maraston2005}
{Maraston} C., 2005, \mnras, 362, 799

\bibitem[{{Merloni} {et~al}\mbox{.}(2003){Merloni}, {Heinz}, \& {di
  Matteo}}]{Merloni2003}
{Merloni} A., {Heinz} S., {di Matteo} T., 2003, \mnras, 345, 1057

\bibitem[{{Mirabel} {et~al}\mbox{.}(1990){Mirabel}, {Booth}, {Johansson},
  {Garay}, \& {Sanders}}]{Mirabel1990}
{Mirabel} I.~F., {Booth} R.~S., {Johansson} L.~E.~B., {Garay} G., {Sanders}
  D.~B., 1990, \aap, 236, 327

\bibitem[{{Mirabel} {et~al}\mbox{.}(1998){Mirabel}, {Vigroux}, {Charmandaris},
  {Sauvage}, {Gallais}, {Tran}, {Cesarsky}, {Madden}, \& {Duc}}]{Mirabel1998}
{Mirabel} I.~F. {et~al.}, 1998, \aap, 333, L1

\bibitem[{{Monreal-Ibero} {et~al}\mbox{.}(2010){Monreal-Ibero}, {Arribas},
  {Colina}, {Rodr{\'{\i}}guez-Zaur{\'{\i}}n}, {Alonso-Herrero}, \&
  {Garc{\'{\i}}a-Mar{\'{\i}}n}}]{MonrealIbero2010}
{Monreal-Ibero} A., {Arribas} S., {Colina} L., {Rodr{\'{\i}}guez-Zaur{\'{\i}}n}
  J., {Alonso-Herrero} A., {Garc{\'{\i}}a-Mar{\'{\i}}n} M., 2010, \aap, 517,
  A28

\bibitem[{{Moster} {et~al}\mbox{.}(2013){Moster}, {Naab}, \&
  {White}}]{Moster2013}
{Moster} B.~P., {Naab} T., {White} S.~D.~M., 2013, \mnras, 428, 3121

\bibitem[{{Mouri}(1994)}]{Mouri1994}
{Mouri} H., 1994, \apj, 427, 777

\bibitem[{{O'Donoghue} {et~al}\mbox{.}(2006){O'Donoghue}, {Buckley}, {Balona},
  {Bester}, {Botha}, {Brink}, {Carter}, {Charles}, {Christians}, {Ebrahim},
  {Emmerich}, {Esterhuyse}, {Evans}, {Fourie}, {Fourie}, {Gajjar}, {Gordon},
  {Gumede}, {de Kock}, {Koeslag}, {Koorts}, {Kriel}, {Marang}, {Meiring},
  {Menzies}, {Menzies}, {Metcalfe}, {Meyer}, {Nel}, {O'Connor}, {Osman}, {Du
  Plessis}, {Rall}, {Riddick}, {Romero-Colmenero}, {Potter}, {Sass},
  {Schalekamp}, {Sessions}, {Siyengo}, {Sopela}, {Steyn}, {Stoffels},
  {Scholtz}, {Swart}, {Swat}, {Swiegers}, {Tiheli}, {Vaisanen}, {Whittaker}, \&
  {van Wyk}}]{DOD2006}
{O'Donoghue} D. {et~al.}, 2006, \mnras, 372, 151

\bibitem[{{Pereira-Santaella} {et~al}\mbox{.}(2015){Pereira-Santaella},
  {Alonso-Herrero}, {Colina}, {Miralles-Caballero}, {P{\'e}rez-Gonz{\'a}lez},
  {Arribas}, {Bellocchi}, {Cazzoli}, {D{\'{\i}}az-Santos}, \& {Piqueras
  L{\'o}pez}}]{PereiraSantaella2015}
{Pereira-Santaella} M. {et~al.}, 2015, \aap, 577, A78

\bibitem[{{Perez} {et~al}\mbox{.}(2011){Perez}, {Michel-Dansac}, \&
  {Tissera}}]{Perez2011}
{Perez} J., {Michel-Dansac} L., {Tissera} P.~B., 2011, \mnras, 417, 580

\bibitem[{{P{\'e}rez-Gonz{\'a}lez}
  {et~al}\mbox{.}(2005){P{\'e}rez-Gonz{\'a}lez}, {Rieke}, {Egami},
  {Alonso-Herrero}, {Dole}, {Papovich}, {Blaylock}, {Jones}, {Rieke}, {Rigby},
  {Barmby}, {Fazio}, {Huang}, \& {Martin}}]{PerezGonzalez2005}
{P{\'e}rez-Gonz{\'a}lez} P.~G. {et~al.}, 2005, \apj, 630, 82

\bibitem[{{Pilkington} {et~al}\mbox{.}(2012){Pilkington}, {Few}, {Gibson},
  {Calura}, {Michel-Dansac}, {Thacker}, {Moll{\'a}}, {Matteucci}, {Rahimi},
  {Kawata}, {Kobayashi}, {Brook}, {Stinson}, {Couchman}, {Bailin}, \&
  {Wadsley}}]{Pilkington2012}
{Pilkington} K. {et~al.}, 2012, \aap, 540, A56

\bibitem[{{Puxley} {et~al}\mbox{.}(1997){Puxley}, {Doyon}, \&
  {Ward}}]{Puxley1997}
{Puxley} P.~J., {Doyon} R., {Ward} M.~J., 1997, \apj, 476, 120

\bibitem[{{Randriamanakoto} {et~al}\mbox{.}(2013){Randriamanakoto},
  {V{\"a}is{\"a}nen}, {Ryder}, {Kankare}, {Kotilainen}, \&
  {Mattila}}]{Randriamanakoto2013}
{Randriamanakoto} Z., {V{\"a}is{\"a}nen} P., {Ryder} S., {Kankare} E.,
  {Kotilainen} J., {Mattila} S., 2013, \mnras, 431, 554

\bibitem[{{Renaud} {et~al}\mbox{.}(2015){Renaud}, {Bournaud}, \&
  {Duc}}]{Renaud2015}
{Renaud} F., {Bournaud} F., {Duc} P.-A., 2015, \mnras, 446, 2038

\bibitem[{{Reunanen} {et~al}\mbox{.}(2002){Reunanen}, {Kotilainen}, \&
  {Prieto}}]{Reunanen2002}
{Reunanen} J., {Kotilainen} J.~K., {Prieto} M.~A., 2002, \mnras, 331, 154

\bibitem[{{Reunanen} {et~al}\mbox{.}(2007){Reunanen}, {Tacconi-Garman}, \&
  {Ivanov}}]{Reunanen2007}
{Reunanen} J., {Tacconi-Garman} L.~E., {Ivanov} V.~D., 2007, \mnras, 382, 951

\bibitem[{{Rich} {et~al}\mbox{.}(2014){Rich}, {Kewley}, \& {Dopita}}]{Rich2014}
{Rich} J.~A., {Kewley} L.~J., {Dopita} M.~A., 2014, \apjl, 781, L12

\bibitem[{{Rich} {et~al}\mbox{.}(2012){Rich}, {Torrey}, {Kewley}, {Dopita}, \&
  {Rupke}}]{Rich2012}
{Rich} J.~A., {Torrey} P., {Kewley} L.~J., {Dopita} M.~A., {Rupke} D.~S.~N.,
  2012, \apj, 753, 5

\bibitem[{{Riffel} {et~al}\mbox{.}(2008){Riffel}, {Pastoriza},
  {Rodr{\'{\i}}guez-Ardila}, \& {Maraston}}]{Riffel2008}
{Riffel} R., {Pastoriza} M.~G., {Rodr{\'{\i}}guez-Ardila} A., {Maraston} C.,
  2008, \mnras, 388, 803

\bibitem[{{Riffel} {et~al}\mbox{.}(2013){Riffel}, {Rodr{\'{\i}}guez-Ardila},
  {Aleman}, {Brotherton}, {Pastoriza}, {Bonatto}, \& {Dors}}]{Riffel2013}
{Riffel} R., {Rodr{\'{\i}}guez-Ardila} A., {Aleman} I., {Brotherton} M.~S.,
  {Pastoriza} M.~G., {Bonatto} C., {Dors} O.~L., 2013, \mnras, 430, 2002

\bibitem[{{Ryder} {et~al}\mbox{.}(2001){Ryder}, {Knapen}, \&
  {Takamiya}}]{Ryder2001}
{Ryder} S.~D., {Knapen} J.~H., {Takamiya} M., 2001, \mnras, 323, 663

\bibitem[{{S{\'a}nchez-Bl{\'a}zquez}
  {et~al}\mbox{.}(2014){S{\'a}nchez-Bl{\'a}zquez}, {Rosales-Ortega},
  {M{\'e}ndez-Abreu}, {P{\'e}rez}, {S{\'a}nchez}, {Zibetti}, {Aguerri},
  {Bland-Hawthorn}, {Catal{\'a}n-Torrecilla}, {Cid Fernandes}, {de Amorim}, {de
  Lorenzo-Caceres}, {Falc{\'o}n-Barroso}, {Galazzi}, {Garc{\'{\i}}a Benito},
  {Gil de Paz}, {Gonz{\'a}lez Delgado}, {Husemann}, {Iglesias-P{\'a}ramo},
  {Jungwiert}, {Marino}, {M{\'a}rquez}, {Mast}, {Mendoza}, {Moll{\'a}},
  {Papaderos}, {Ruiz-Lara}, {van de Ven}, {Walcher}, \&
  {Wisotzki}}]{Sanchez2014}
{S{\'a}nchez-Bl{\'a}zquez} P. {et~al.}, 2014, \aap, 570, A6

\bibitem[{{Sanders} \& {Mirabel}(1996)}]{Sanders1996}
{Sanders} D.~B., {Mirabel} I.~F., 1996, \araa, 34, 749

\bibitem[{{Shapiro} {et~al}\mbox{.}(2008){Shapiro}, {Genzel}, {F{\"o}rster
  Schreiber}, {Tacconi}, {Bouch{\'e}}, {Cresci}, {Davies}, {Eisenhauer},
  {Johansson}, {Krajnovi{\'c}}, {Lutz}, {Naab}, {Arimoto}, {Arribas},
  {Cimatti}, {Colina}, {Daddi}, {Daigle}, {Erb}, {Hernandez}, {Kong},
  {Mignoli}, {Onodera}, {Renzini}, {Shapley}, \& {Steidel}}]{Shapiro2008}
{Shapiro} K.~L. {et~al.}, 2008, \apj, 682, 231

\bibitem[{{Shibuya} {et~al}\mbox{.}(2016){Shibuya}, {Ouchi}, {Kubo}, \&
  {Harikane}}]{Shibuya2016}
{Shibuya} T., {Ouchi} M., {Kubo} M., {Harikane} Y., 2016, \apj, 821, 72

\bibitem[{{Smith} \& {Hancock}(2009)}]{Smith2009}
{Smith} B.~J., {Hancock} M., 2009, \aj, 138, 130

\bibitem[{{Tekola} {et~al}\mbox{.}(2014){Tekola}, {Berlind}, \&
  {V{\"a}is{\"a}nen}}]{Tekola2014}
{Tekola} A.~G., {Berlind} A.~A., {V{\"a}is{\"a}nen} P., 2014, \mnras, 439, 3033

\bibitem[{{Toomre}(1964)}]{toomre}
{Toomre} A., 1964, \apj, 139, 1217

\bibitem[{{V{\"a}is{\"a}nen} {et~al}\mbox{.}(2014){V{\"a}is{\"a}nen}, {Barway},
  \& {Randriamanakoto}}]{Vaisanen2014}
{V{\"a}is{\"a}nen} P., {Barway} S., {Randriamanakoto} Z., 2014, \apjl, 797, L16

\bibitem[{{V{\"a}is{\"a}nen} {et~al}\mbox{.}(2008){V{\"a}is{\"a}nen},
  {Mattila}, {Kniazev}, {Adamo}, {Efstathiou}, {Farrah}, {Johansson},
  {{\"O}stlin}, {Buckley}, {Burgh}, {Still}, \& {Zijlstra}}]{Vaisanen2008}
{V{\"a}is{\"a}nen} P. {et~al.}, 2008, \mnras, 384, 886

\bibitem[{{V{\"a}is{\"a}nen} {et~al}\mbox{.}(2012){V{\"a}is{\"a}nen},
  {Rajpaul}, {Zijlstra}, {Reunanen}, \& {Kotilainen}}]{Vaisanen2012}
{V{\"a}is{\"a}nen} P., {Rajpaul} V., {Zijlstra} A.~A., {Reunanen} J.,
  {Kotilainen} J., 2012, \mnras, 420, 2209

\bibitem[{{Westmoquette} {et~al}\mbox{.}(2012){Westmoquette}, {Clements},
  {Bendo}, \& {Khan}}]{Westmoquette2012}
{Westmoquette} M.~S., {Clements} D.~L., {Bendo} G.~J., {Khan} S.~A., 2012,
  \mnras, 424, 416

\bibitem[{{Whitaker} {et~al}\mbox{.}(2012){Whitaker}, {van Dokkum}, {Brammer},
  \& {Franx}}]{Whitaker2012}
{Whitaker} K.~E., {van Dokkum} P.~G., {Brammer} G., {Franx} M., 2012, \apjl,
  754, L29

\bibitem[{{Winge} {et~al}\mbox{.}(2009){Winge}, {Riffel}, \&
  {Storchi-Bergmann}}]{Winge2009}
{Winge} C., {Riffel} R.~A., {Storchi-Bergmann} T., 2009, \apjs, 185, 186

\bibitem[{{Yuan} {et~al}\mbox{.}(2010){Yuan}, {Kewley}, \&
  {Sanders}}]{Yuan2010}
{Yuan} T.-T., {Kewley} L.~J., {Sanders} D.~B., 2010, \apj, 709, 884

\end{thebibliography}

\end{document}